\documentclass[a4paper,11pt]{article}
\pdfoutput=1 % if your are submitting a pdflatex (i.e. if you have
\usepackage{jheppub} % for details on the use of the package, please
\usepackage[T1]{fontenc} % if needed
\usepackage{footnote}

\title{\boldmath Baryogenesis from ultra-slow-roll inflation}

\author[a]{Yi-Peng Wu,}
\author[a, b, c]{Elena Pinetti,}
\author[a, d]{Kalliopi Petraki,}
\author[e, f, g]{and Joseph Silk}

\affiliation[a]{Laboratoire de Physique Th\'{e}orique et Hautes Energies (LPTHE), \\
	UMR 7589 CNRS \& Sorbonne Universit\'{e}, 4 Place Jussieu, F-75252, Paris, France}
\affiliation[b]{Dipartimento di Fisica, Università di Torino \& INFN, Sezione di Torino, via P. Giuria 1, I-10125 Torino, Italy}
\affiliation[c]{Theoretical Astrophysics Department, Fermi National Accelerator Laboratory, Batavia, Illinois, 60510, USA}
\affiliation[d]{Nikhef, 
	Science Park 105, 1098 XG Amsterdam, The Netherlands}
\affiliation[e]{Institut d’Astrophysique de Paris, UMR 7095 CNRS \& Sorbonne Universit\'{e}, 98 bis boulevard Arago, F-75014 Paris, France}
\affiliation[f]{Department of Physics and Astronomy, The Johns Hopkins University, 3400 N. Charles
	Street, Baltimore, MD 21218, U.S.A.}
\affiliation[g]{Beecroft Institute for  Particle Astrophysics and Cosmology, University of Oxford, Keble
	Road, Oxford OX1 3RH, U.K.}

\emailAdd{ywu@lpthe.jussieu.fr}
\emailAdd{elena.pinetti@unito.it}
\emailAdd{kpetraki@lpthe.jussieu.fr}
\emailAdd{silk@iap.fr}

\abstract{ The ultra-slow-roll (USR) inflation represents a class of single-field models with sharp deceleration of the rolling dynamics on small scales, leading to a significantly enhanced power spectrum of the curvature perturbations and primordial black hole (PBH) formation. Such a sharp transition of the inflationary background can trigger the coherent motion of scalar condensates with effective potentials governed by the rolling rate of the inflaton field. We show that a scalar condensate carrying (a combination of) baryon or lepton number can achieve successful baryogenesis through the Affleck-Dine mechanism from unconventional initial conditions excited by the USR transition. Viable parameter space for creating the correct baryon asymmetry of the Universe naturally incorporates the specific limit for PBHs to contribute significantly to  dark matter, shedding light on the cosmic coincidence problem between the baryon and dark matter densities today.}

\begin{document} 
\maketitle
\flushbottom

\section{Introduction}
\label{Sec:intro}
Cosmic inflation naturally gives rise to coherent production of scalar condensates with extremely large vacuum expectation values (VEVs). These VEVs are characterized by the Hubble scale during inflation, $H_\ast$, which could be around the order of $10^{13}$ GeV \cite{Tristram:2020wbi,Akrami:2018odb}. If some of these scalar condensates carry baryon or lepton numbers from the Standard Model or dark sectors of the particle theories, their relaxation and decay after inflation ends can create a non-negligible baryon (lepton) asymmetry to the Universe. This is usually referred to as the Affleck-Dine (AD) mechanism of baryogenesis \cite{Affleck:1984fy,Linde:1985gh,Dolgov:1991fr,Dine:2003ax}.

In order to obtain a quantitative prediction about the amount of baryon asymmetry produced by the AD mechanism from the energy scale of inflation, there are at least two essential questions requiring clarification at the beginning of the process, as pointed out in \cite{Dine:1995kz}: (1) What are the origins of the baryon/lepton number violating interactions in the scalar potential, and (2) what are the initial conditions for the relaxation of scalar condensates?   

So far, the mainstream research on AD baryogenesis has been devoted to the presence of ``flat directions'' in the early Universe arising from, typically, the supersymmetric extension of the Standard Model field theories \cite{Dine:1995kz}. The necessary supersymmetry breaking at the energy scale of inflation provides answers to both questions (1) and (2). Firstly, self-couplings of flat directions in the Kahler potential (or the superpotential) generically foster baryon/lepton number (B/L) violating terms that are non-renormalizable. These B/L violating terms can have complex coefficients to serve as the source of CP violation in the AD mechanism, yet this is not a necessary condition since CP violation can emerge from the stochastic pick-up of scalar VEVs during inflation \cite{Dine:1995kz,Wu:2020pej,Hook:2015foa}. Secondly, flat directions are in general lifted by soft supersymmetry violating terms and higher-order non-renormalizable terms in the superpotential with parameters of order $H_\ast$. These parameters determine the global minimum of the effective potential (and thus the VEVs of flat directions). Thanks to the tachyonic instability at the potential origin caused by soft terms, initial VEVs for the AD mechanism may be larger than $H_\ast$ by $10^2$ $-$ $10^3$ times, ensuring  successful baryogenesis over a wide parameter space.  

In this work, we explore a novel class of initial conditions for AD baryogenesis in the unusual limit with VEVs of the charged scalar (the AD field) provided that they are much smaller than $H_\ast$. Our scenario does not rely on a tachyonic mass term, so that the average value of the AD field decays exponentially in time. This is similar to that of the flat direction scenario with a minimal Kahler potential in which case the AD mechanism is expected to fail~\cite{Dine:1995kz}. The key difference with the mainstream scenario is that the AD field is dynamically excited around a later phase of inflation so that the scalar condensate is not in an equilibrium state \cite{Starobinsky:1994bd}. With a slow but non-zero coherent motion of the AD field by the end of inflation, we show that initial VEVs around $10^{-2}$ - $10^{-1}$ in  units of $H_\ast$ are sufficient to produce the baryon asymmetry observed today.

As a concrete demonstration of the basic idea, we consider that the background spacetime of inflation is created by single-field models that exhibit multi-stage transitions of the rolling dynamics of the inflaton field \cite{Yokoyama:1998pt,Saito:2008em,Garcia-Bellido:2017mdw,Germani:2017bcs,Motohashi:2017kbs,Cicoli:2018asa,Ozsoy:2018flq,Byrnes:2018txb,Liu:2020oqe,Cheng:2018qof,Ozsoy:2019lyy,Ballesteros:2020sre,Ng:2021hll,Carrilho:2019oqg,Kannike:2017bxn}. This class of models is proposed for primordial black hole (PBH) formation on small scales with significantly enhanced curvature perturbation induced by a sharp deceleration of inflaton (from the primary slow-roll phase on large scales). The so-called ultra-slow-roll (USR) inflation \cite{Kinney:2005vj,Tsamis:2003px,Martin:2012pe,Motohashi:2014ppa,Anguelova:2017djf} is a special case of  interest, which in particular provides a convenient limit for illustration.

We show that the presence of derivatives coupling between the inflaton and the AD field in general introduces soft effective mass terms governed by the rolling, and more importantly, the rate of rolling of inflaton. Due to the sharp transition of the inflaton rolling rate in the USR scenario for PBH formation, coherent motion of the AD field condensate can be triggered during inflation, leading to  successful baryogenesis from initial conditions in the parameter space that were usually omitted in previous investigations.

There is a crucial asset in considering baryogenesis triggered by inflation scenarios of the USR class, since PBHs are currently under close examination of their feasibility for contributing all (or a significant fraction) of dark matter. Suppose that PBHs occupy more than 10 percent of the dark matter density, their energy density today is of the same order as that of the baryons. This could be considered as a special version of the ``cosmic coincidence problem.'' 

Baryogenesis initiated by USR inflation is a mechanism in parallel with that of PBH formation, and can be realized even without the presence of any PBHs today. Thus it is not identical to the idea where baryon asymmetry of the Universe is created by existing PBHs \cite{Turner:1979bt,Barrow:1990he,Majumdar:1995yr,Baumann:2007yr,Bambi:2008hp,Hamada:2016jnq,Hooper:2020otu,Ambrosone:2021lsx,JyotiDas:2021shi}, providing a different insight to the cosmic coincidence between dark matter and baryon densities sourced by PBHs  \cite{Fujita:2014hha,Garcia-Bellido:2019vlf} or originated from the asymmetric dark sector in particle physics \cite{Petraki:2013wwa,vonHarling:2012yn,Bell:2011tn,Flores:2020drq}.
Based on the assumption of PBH dark matter, the scenario investigated in this work allows a more systematic study on the viable parameter space for the cosmic coincidence problem as model parameters of the USR inflation simultaneously control the dark matter density and the baryon asymmetry of the Universe.

This work is organized as follows: in Section~\ref{Sec:inflation_PBH}, we outline the structure of the inflaton dynamics in the (quasi) USR limit, the template of the corresponding power spectrum of the curvature perturbation, and the coherent motion of a massive scalar field under such an inflationary background. In Section~\ref{Sec. Baryogenesis_during_inflation}, we provide a simple realization of the AD field with effective mass terms controlled by parameters of the USR inflation. We then compute the AD field evolution during inflation and estimate the amount of baryon asymmetry created at the end of inflation. These results are initial conditions for Section~\ref{Sec. baryogenesis_after_inflation} to obtain the final baryogenesis in the  radiation-dominated era  of the Universe. Finally, in Section~\ref{Sec.B_PBH_correlation} we take into account the associated PBH abundance in terms of the USR parameters based on several statistical approaches that address different uncertainties about the PBH formation from inflation. We explore the viable parameter space for PBHs to be a significant contributor to  dark matter with the correct amount of baryon asymmetry. Conclusions for baryogenesis and implications for the cosmic coincidence problem are given in Section~\ref{Sec:conclusion}.

\section{Inflation and primordial black holes}\label{Sec:inflation_PBH}
Models of inflation investigated in this work are motivated by primordial black hole (PBH) production via an enhanced power spectrum $P_\zeta$ of the curvature perturbation $\zeta$ on small scales. In single-field inflationary framework, the enhancement of $P_\zeta$ can be realized by a sharp deceleration of the inflaton field, $\phi$, in its potential $V(\phi)$ that causes the decrease of the first slow-roll parameter $\epsilon_H \equiv -\dot{H}/H^2 \approx \epsilon_V$ by many orders of magnitude, where $\epsilon_V \equiv (V_{,\phi}/V)^2M_P^2$ denotes the first slow-roll parameter for the inflaton potential. Throughout this work, we will keep in mind the key parameter that controls such a deceleration, defined as
\begin{equation}\label{def. delta}
\delta = \frac{\ddot{\phi}}{H\dot{\phi}},
\end{equation} 
where $\delta \approx \epsilon_V - \eta_V$ and $\eta_V \equiv V_{,\phi\phi}/VM_P^2$ denotes the second slow-roll parameter for the potential. $\delta$ can be $\mathcal{O}(1)$ or larger due to the significant violation of the slow-roll conditions from $\eta_V$. We restrict ourselves to models that satisfy the condition $\epsilon_H \approx \epsilon_V \ll 1$ where the background is well described by the de Sitter space.

\subsection{Ultra-slow-roll (USR) templates}\label{Sec. USR_template}
\begin{figure}
	\begin{center}
		\includegraphics[width=7cm]{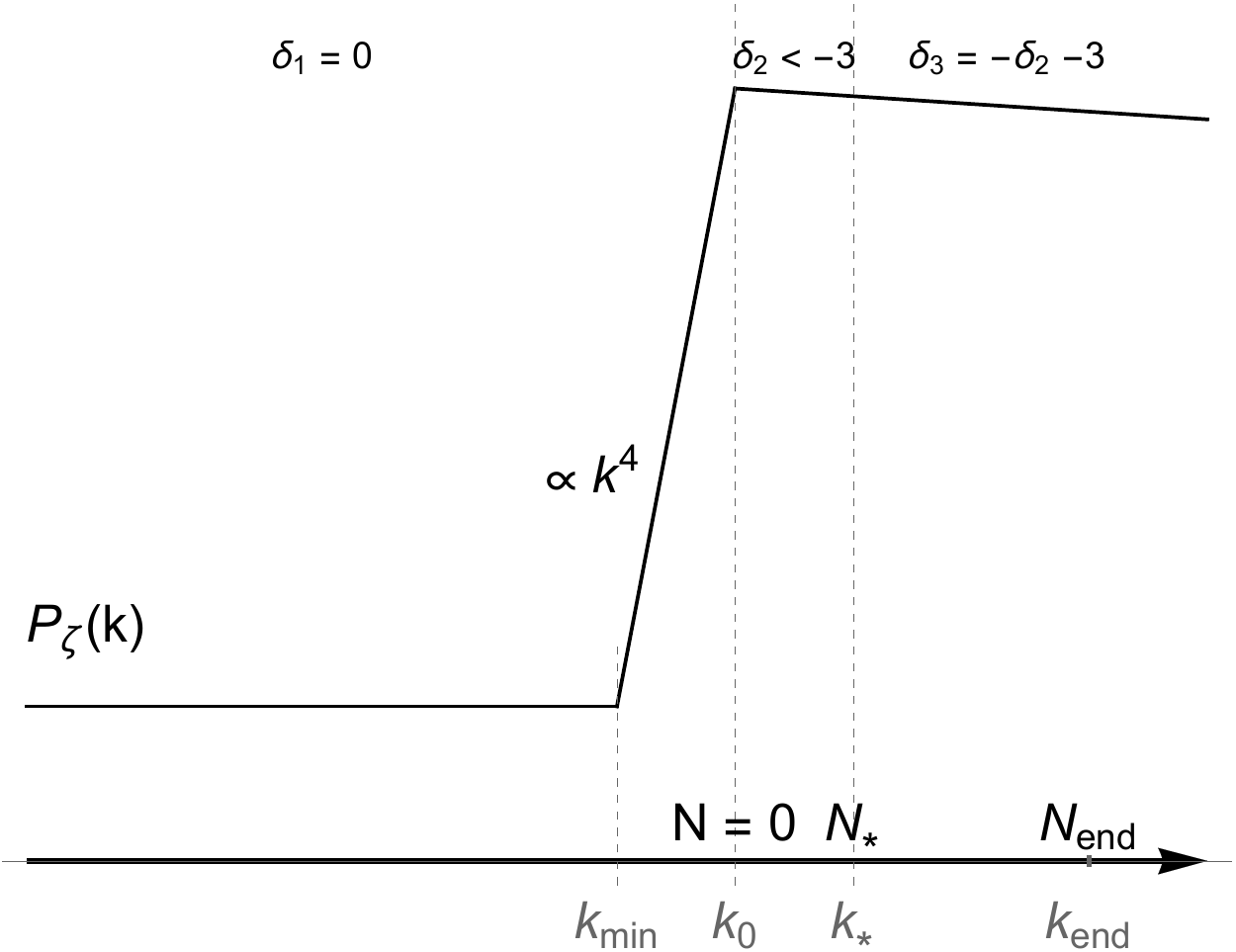} 
%\par
	\end{center}
	\caption{Analytic template for the power spectrum of the curvature perturbation for a 3-stage constant-rate inflation with Phase 1 $\delta_1 = 0$ of the primary slow-roll and Phase 2 $\delta_2 = -3.05$ in the (quasi-)ultra-slow-roll limit.  \label{fig:USR_spectrum}}
\end{figure}

To illustrate the basic idea for baryogenesis, we will assume that inflation is composed of three essential phases with a nearly constant rate-of-rolling ($\delta \approx$ const.) during each period. The analytic structure of $P_\zeta$ with the constant-rate condition $\dot{\delta} = 0$ has been well studied in multi-stage inflation  \cite{Byrnes:2018txb,Liu:2020oqe,Cheng:2018qof,Ozsoy:2019lyy,Ballesteros:2020sre,Ng:2021hll,Carrilho:2019oqg}.
 The first stage of inflation is the primary slow-roll phase with a nearly scale-invariant power spectrum probed by cosmic microwave background (CMB) observations and we take $\delta_1 = \delta (t < t_0) \rightarrow 0$ for simplicity. Superhorizon enhancement of $P_\zeta$ occurs when the second stage ($t_0 \leq t < t_\ast$) has a constant rate $\delta_2 < -3/2$ \cite{Liu:2020oqe}. Except for further notification, in the following discussion we focus on a second stage of inflation close to (but not exactly in) the ultra-slow-roll (USR) limit \cite{Kinney:2005vj,Tsamis:2003px,Martin:2012pe,Motohashi:2014ppa,Anguelova:2017djf}, where  
\begin{equation}\label{USR_delta_definition}
\delta_2 = \frac{\ddot{\phi}}{H\dot{\phi}} \lesssim -3, \quad t_{0} < t < t_{\ast}.
\end{equation} 
To suppress the non-Gaussian tail in the high-sigma limit of $\zeta$ led by quantum diffusion in the exact USR limit ($\delta_2 \rightarrow -3$) \cite{Biagetti:2018pjj,Ezquiaga:2019ftu,Pattison:2021oen,Figueroa:2020jkf,Biagetti:2021eep}, we put an upper bound to $\delta_2$ as discussed in Appendix~\ref{Appd_inflaton_mass}.
Note that for $\delta \leq -3$, the curvature perturbation $\zeta$ is led by the late-time scaling dimension (or conformal weight) $\Delta \equiv 3/2 - \sqrt{9/4 +3\delta + \delta^2} = 3/2 - \vert 3/2 + \delta\vert$, and thus the dimensionless power spectrum $P_\zeta \sim k^{2\Delta} \rightarrow k^0$ as $\delta \rightarrow -3$. The USR limit is therefore a specific case in the slow-roll violation regime that can give rise to a scale-invariant power spectrum as that from the standard slow-roll inflation ($\delta \rightarrow 0$).

The momentum scaling of the power spectrum $P_\zeta(k)$ from a 3-stage constant rate inflation can be summarized as \cite{Liu:2020oqe,Ng:2021hll}:
\begin{eqnarray}\label{CR_template}
P_\zeta = \left\{
\begin{array}{ll}
A_{\rm CMB} & k < k_{\rm min}, \\
A_{\rm PBH} (k/k_0)^4, & k_{\rm min} < k < k_0, \\
A_{\rm PBH} (k/k_0)^{6 + 2\delta_2}, & k > k_0,
\end{array}
\right.
\end{eqnarray}
where $A_{\rm CMB} \approx H_\ast^2/(\epsilon_{\rm CMB} M_P^2)$ measured on CMB scales has negligible contribution for PBH formations, and the enhanced spectral amplitude
\begin{equation}\label{A_zeta_analytic}
A_{\rm PBH} \approx  \frac{H_\ast^2}{\epsilon_{\rm CMB} M_P^2} \left(\frac{k_{0}}{k_{\ast}}\right)^{6 + 4 \delta_2}.
\end{equation}
Here $M_P$ is the reduced Planck mass and $H_\ast$ is the Hubble parameter during inflation and $0 \leq \epsilon_{\rm CMB} <  0.0063$ \cite{Akrami:2018odb}. 
In Figure~\ref{fig:USR_spectrum} we plot the case in the USR limit with $\delta_2 = -3$. We choose $N = \ln a = 0$ at the time $t =t_0$ where $k_0 = a(t_0)H_\ast$ crosses the horizon so that $A_{\rm PBH}/A_{\rm CMB} = e^{-N_\ast(6+4\delta_2)}$. Models of inflation that realize the power spectrum given in Figure~\ref{fig:USR_spectrum} are sometimes called ``punctuated inflation.'' 
\cite{Ragavendra:2020sop,Jain:2008dw,Allahverdi:2006we,Jain:2009pm}
See also \cite{Byrnes:2018txb,Biagetti:2018pjj,Motohashi:2019rhu,Ragavendra:2020sop} for PBHs from inflationary scenarios that exhibit a secondary slow-roll phase.

The transition from the primary slow-roll phase (Phase 1) to the USR phase (Phase 2) shows an apparent violation of the continuity of the time derivative for the leading $\zeta$ mode with $\Delta_1 = 0$ on superhorizon scales. This is due to the efficient entropy production driven by the sharp deceleration of inflaton that violates the adiabaticity of the curvature perturbation \cite{Ragavendra:2020sop,Ng:2021hll,Leach:2000yw}. Subleading entropy modes with the next-to-lowest scaling dimension come to dominate the power spectrum in the range of $k_{\rm min} \leq k < k_0$ with the special power $P_\zeta \sim k^4$, where $k_{\rm min } \approx k_0 (k_0/k_\ast)^{3/2}$ for $\delta_2 = -3$.\footnote{The leading scaling dimension for the curvature perturbation $\zeta$ in slow-roll is the same as in USR ($\Delta_{\rm SR} = \Delta_{\rm USR} = 0$). However, the leading scaling dimension for the time derivative $\dot{\zeta}$ in USR is $\Delta_{\rm USR} -1$, yet it is $\Delta_{\rm SR} +1$ in slow-roll. Thus the steepest growth $P_\zeta \sim k^4$ occurs only for the slow-roll $\rightarrow$ USR transition \cite{Byrnes:2018txb} and it does not occur for the inverse USR $\rightarrow$ slow-roll transition \cite{Cai:2017bxr}. For a more detailed discussion, see \cite{Ng:2021hll}. }
 The $k^4$ scaling of $P_\zeta$ is firstly identified in USR inflation by Leach-Sasaki-Wands-Liddle \cite{Leach:2001zf}, and this mechanism is also called as the steepest growth of the power spectrum \cite{Byrnes:2018txb,Carrilho:2019oqg}.

The transition from the USR phase to the final phase (Phase 3) with a non-negative rate $\delta_3 \geq 0$ is necessary for stopping the growth of the power spectrum. For inflation models that can realized a stable negative rate, $\delta_2 \leq -3$ with $\dot{\delta}_2 \approx 0$, in Phase 2, the value of $\delta_3$ is constrained by the continuity of the scaling dimension $\Delta$, where the condition $\Delta_2 = \Delta_3 = -\delta_3$ indicates that $\delta_3 = -\delta_2 -3$. This is due to the non-violation of the adiabatic condition in the acceleration phase of inflaton, and the effective mass of inflaton retains continuous \cite{Ng:2021hll}. Thus if Phase 2 goes to the USR limit with $\delta_2\rightarrow -3$, one finds that $\delta_3 \rightarrow 0$ where Phase 3 must be a secondary slow-roll phase.

The termination of the secondary slow-roll phase in this class of models relies an additional assumptions \cite{Ragavendra:2020sop,Jain:2008dw,Allahverdi:2006we,Jain:2009pm}. We will consider that inflation is terminated by (1) the sudden decay of inflaton into radiation at $N_{\rm end}$ and drives instantaneous reheating (Section~\ref{Sec. Baryogenesis_during_inflation}) or (2) the coherent oscillations of the inflaton as it rolls into a deep valley and obtains a large mass (Section~\ref{Sec. baryogenesis_after_inflation}). 

\subsection{Massive scalars under USR transition}\label{Sec. massive_scalar_USR}
Before going to the concrete scenario for baryogenesis, we derive in this section some useful formulae that describe the coherent motion of massive scalar fields during the multi-stage constant-rate inflation illustrated above, focusing on the limit where Phase 2 is USR. We are interested in a scenario where a spectator field, $\sigma$, with negligible energy density during inflation receives different effective masses in each phase, where $m_\sigma = m_i$ for $i = 1, 2, 3$ in Phase $i$. The vacuum expectation value (VEV) in Phase 1 is developed naturally by the stochastic effect in equilibrium \cite{Starobinsky:1994bd}. This gives the initial conditions for Phase 2 as
\begin{align}\label{initial_condition_Phase1}
\sigma_1(t_0) = \sqrt{\frac{3}{8\pi^2}} \frac{H_\ast^2}{m_1}, \quad \dot{\sigma}_1(t_0) = 0,
\end{align}  
where the VEV is the result of a massive non-interacting scalar.

With the transition from the primary slow-roll phase to USR, $\sigma$ gains a different mass $m_2 \neq m_1$ which suddenly drives the coherent condensate out of equilibrium in the potential. This leads to the evolution of the VEV with respect to a simplified equation of motion as
\begin{align}
\ddot{\sigma}_2 + 3 H_\ast \dot{\sigma}_2 + m_2^2 \,\sigma_2 = 0, \qquad t_0\leq t < t_\ast.
\end{align}
According to the initial conditions given by \eqref{initial_condition_Phase1}, the solution of $\sigma_2$ reads \cite{Antoniadis:2011ib}:
\begin{align}
\sigma_2(N) = \sqrt{\frac{3}{8\pi^2}} \frac{H_\ast^2}{m_1} \frac{1}{2\nu_2}
\left(\Delta_2^+ e^{-\Delta_2^- N} - \Delta_2^- e^{-\Delta_2^+ N}\right),
\end{align}
where we use $N = H_\ast t$ with $N_0 = H_\ast t_0 =0$, and the scaling dimensions (conformal weights) for a massive scalar in the de Sitter space \cite{Strominger:2001pn,Ng:2012xp,Jafferis:2013qia,Antoniadis:2011ib} are given by
\begin{align}
\Delta_2^\pm \equiv \frac{3}{2} \pm \nu_2, \qquad \nu_2 = \sqrt{\frac{9}{4} - \frac{m_2^2}{H_\ast^2}}.
\end{align}
Note that this is the leading scaling dimension for the coherent motion of $\sigma$ (to be distinguished from those for the curvature perturbation $\zeta$ in Section~\ref{Sec. USR_template}). With $\Delta_2^-\Delta_2^+ = 9/4 -\nu_2^2 = m_2^2/H_\ast^2$, the time-derivative of $\sigma_2$ reads
\begin{align}
\frac{d\sigma_2}{d N} = \frac{\dot{\sigma}_2}{H_\ast} = \sqrt{\frac{3}{8\pi^2}} \frac{m_2^2}{m_1} \frac{1}{2\nu_2}
\left( e^{-\Delta_2^+ N} -  e^{-\Delta_2^- N}\right),
\end{align}
which gives non-zero initial conditions for Phase 3 at $N = N_\ast$.

After the scalar condensate is excited by the slow-roll to USR transition, the equation of motion for $\sigma$ from USR to the secondary slow-roll phase (Phase 3) is unchanged but the non-zero time-derivative in Phase 2 asks the solution to take the general form as
\begin{align}
\sigma_3(N) = C e^{-\Delta_3^+ N} + D e^{-\Delta_3^- N},
\end{align}
where similarly, the leading scaling dimension is fixed by the mass in Phase 3 as
\begin{align}\label{Phase3_Delta}
\Delta_3^\pm \equiv \frac{3}{2} \pm \nu_3, \qquad \nu_3 = \sqrt{\frac{9}{4} - \frac{m_3^2}{H_\ast^2}}.
\end{align}
Matching boundary conditions at the time slice $N = N_\ast$ at the end of USR, one finds
\begin{align}\label{Phase3_coefficient}
C = \sqrt{\frac{3}{8\pi^2}} \frac{H_\ast^2}{m_1} &\frac{\Delta_2^-}{\Delta_2^- - \Delta_2^+} 
\frac{e^{\Delta_3^+ N_\ast}}{\Delta_3^- -\Delta_3^+} 
\nonumber\\
&\times 
\left[\left(\Delta_3^- - \Delta_2^+\right) e^{-\Delta_2^+ N_\ast} + \Delta_2^+\left(1-\frac{\Delta_3^-}{\Delta_2^-}\right) e^{-\Delta_2^- N_\ast}\right],
\\
D = \sqrt{\frac{3}{8\pi^2}} \frac{H_\ast^2}{m_1} &\frac{\Delta_2^-}{\Delta_2^- - \Delta_2^+}
\frac{e^{\Delta_3^- N_\ast}}{\Delta_3^+ -\Delta_3^-} 
\nonumber\\
&\times 
\left[\left(\Delta_3^+ - \Delta_2^+\right) e^{-\Delta_2^+ N_\ast} + \Delta_2^+\left(1-\frac{\Delta_3^+}{\Delta_2^-}\right) e^{-\Delta_2^- N_\ast}\right].
\end{align}
For a constant mass $m_i /H_\ast \leq 3/2$, $\nu_i$ is real so that the positive branch $\Delta^+$ decays much faster than the negative branch $\Delta^-$. For $m_i /H_\ast > 3/2$, $\nu_i$ or $\Delta^\pm_i$ contain imaginary parts so that both $\pm$ branches contribute to the late-time dynamics of the mass scalar. In this work, we focus on the mass range $m_i /H_\ast \leq 3/2$ and thus only the negative branch survives at the end of inflation as
\begin{align}\label{late_time_Phase3}
\sigma_3(N_{\rm end}) \rightarrow D e^{-\Delta_3^- N_{\rm end}}, \qquad 
\left.\frac{d\sigma_3}{d N} \right\vert_{N_{\rm end}} \rightarrow -\Delta_3^- D e^{-\Delta_3^- N_{\rm end}}.
\end{align}
The results in \eqref{late_time_Phase3} are the two most relevant quantities for the computation of baryon production from a charged scalar field at or after the end of inflation.

\section{Baryogenesis during inflation}\label{Sec. Baryogenesis_during_inflation}
In most of the cases, baryon or lepton number production during inflation is considered negligible due to the exponential dilution of the particle number density in time. However, for baryon production triggered by the phase transition of USR at $N = N_0$ on scales much smaller than those constrained by CMB observations, the duration $\Delta N = N_{\rm end} - N_0$ is less constrained as long as the total duration $N_{\rm end} - N_{\rm CMB} \gtrsim \mathcal{O}(50)$.  

Baryogenesis investigated in this section is driven by a charged scalar $\sigma$ that possesses a global $U(1)$ particle number and develops non-zero expectation values during inflation. The baryon or lepton number excitation is triggered by the phase transition of the USR dynamics and it is essentially described by the Affleck-Dine (AD) mechanism \cite{Affleck:1984fy}, except for the fact that we evaluate the process in the (quasi-)de Sitter space.

\subsection{Asymmetric scalar masses}

To illustrate how the transition of inflaton rolling rate $\delta$ given by \eqref{def. delta} can lead to a transition of effective masses for a charged scalar $\sigma$ (the AD field), let us consider 
\begin{align}\label{Lagrangian_AD}
\mathcal{L} = \mathcal{L}_\phi +\vert \partial\sigma\vert^2 + m_\sigma^2 \vert \sigma\vert^2 + \Delta\mathcal{L},
\end{align} 
where $\mathcal{L}_\phi $ describes the multi-stage constant-rate (including the USR) model for PBH formation. $\Delta\mathcal{L}$ includes self-interactions of $\sigma$ and derivative couplings with $\phi$ of the form
\begin{align}\label{Lagrangian_coupling}
\Delta\mathcal{L} \supset \frac{c_1}{\Lambda} \left\vert  \sigma^2\right\vert \square \phi 
+ \frac{c_2}{\Lambda}\partial_\mu \phi \left[\sigma\partial^\mu \sigma + \sigma^\ast\partial^\mu \sigma^\ast\right] + 
 \frac{c_3}{\Lambda^2} \left(\partial\phi\right)^2\vert\sigma\vert^2,
\end{align}
where we have imposed C and CP to the system so that $c_1$, $c_2$ and $c_3$ are $\mathcal{O}(1)$ real constants.\footnote{In general, $c_1$, $c_2$ and $c_3$ can have imaginary components given by non-zero phases. However, the combination of $c_1$ and $c_3$ terms with their complex conjugates only gives a redefinition of $c_1$ and $c_3$, while the phase of $c_2$ can be absorbed into $\sigma$ via a constant shift of its phase term.} The derivative couplings used in \eqref{Lagrangian_coupling} are similar to those of the inflaton-induced chemical potential for an enhanced charged scalar production \cite{Wang:2019gbi,Wang:2020ioa,Bodas:2020yho}.
However, CP-violating couplings, such as $\partial_\mu \phi(\sigma^\ast \partial^\mu\sigma - \sigma\partial^\mu \sigma^\ast)$ \cite{Bodas:2020yho} is forbidden for simplicity. The necessary C/CP violation for baryogenesis is spontaneously realized by the initial condensation of the charged scalar at the VEVs during inflation.

One can rewrite the Lagrangian for $\sigma$ with respect to the mass eigenstates $\sigma_{\pm}$, based on the decomposition $\sigma \equiv (\sigma_{-} + i \sigma_{+})/\sqrt{2}$, to obtain a system for two real scalars as
\begin{align}\label{Lagrangian_shift_pm_decouple}
\mathcal{L}_\sigma =& \frac{1}{2}\left(\partial\sigma_{-}\right)^2 + \frac{1}{2} 
\left[m_\sigma^2 + \frac{c_1 - c_2}{\Lambda} \square \phi  + \frac{c_3}{\Lambda^2}\left(\partial\phi\right)^2\right] \sigma_{-}^2 
\nonumber\\
&+ \frac{1}{2}\left(\partial\sigma_{+}\right)^2 + \frac{1}{2} 
\left[m_\sigma^2 + \frac{c_1 + c_2}{\Lambda} \square \phi  + \frac{c_3}{\Lambda^2}\left(\partial\phi\right)^2\right] \sigma_{+}^2.
\end{align}
The mass eigenstates provide a convenient basis for the study of baryon number production, and, unless with a fine tuning of the coefficients, they in general have non-degenerated effective masses
\begin{align}\label{mass_pm}
m_\pm^2 = m_\sigma^2 + \frac{c_1 \pm c_2}{\Lambda} \square\phi + \frac{c_3}{\Lambda^2} \left(\partial\phi\right)^2.
\end{align}
$\Lambda$ is the cutoff scale for the effective field theory that satisfies $H_\ast \ll \Lambda \leq M_P$.

We now apply the multi-stage constant-rate inflation described in Section~\ref{Sec. USR_template} to specify the spatially homogeneous dynamics of $\phi$. With a constant $\delta$, we have $\square\phi = -\ddot{\phi} -3H\dot{\phi} \approx - (\delta + 3)\sqrt{2\epsilon_H} M_PH_\ast^2$ and $(\partial\phi)^2 = \dot{\phi}^2 \approx 2\epsilon_HM_P^2 H_\ast^2$. If $\delta \neq 0$, the first slow-roll parameter $\epsilon_H$ is evolving in each phase where $\epsilon_2 = \epsilon_{\rm CMB}e^{2\delta_2 N}$ in Phase 2 and $\epsilon_3 = \epsilon_\ast e^{2\delta_3(N-N_\ast)}$ in Phase 3 with $\epsilon_\ast = \epsilon_{\rm CMB}e^{2\delta_2 N_\ast}$. As a result, the effective masses for $\sigma_\pm$ change with the rolling rate of inflaton in different phases:
\begin{align}
m_{i\pm}^2 = m_\sigma^2 + \frac{c_1 \pm c_2}{\Lambda}[-(\delta_i + 3)]\sqrt{2\epsilon_i} M_PH_\ast^2
+ \frac{c_3}{\Lambda^2} 2\epsilon_i M_P^2H_\ast^2.
\end{align}
For the USR scenario of our interest ($\delta_1 = 0$, $\delta_2 = -3$ and $\delta_3 = -\delta_2 -3 = 0$), the masses $m_{1\pm}$ and $m_{3\pm}$ in Phase 1 \& 3 are always constants.
The time-dependence of $m_{2\pm}$ replies on the choices of parameters, and there are two common assumptions for the mass spectrum of the AD field:

\bigbreak\noindent
\textbf{Heavy AD field.} Charged scalars arising as the superpartners of the Standard Model fermions are natural candidates for the AD field \cite{Dine:1995kz}. The necessary supersymmetry breaking generically induces soft masses for those charged scalars of order the Hubble constant $H_\ast$ due to inflation. Baryon-violating terms residing in the superpotential, a non-minimal coupling with gravity or the stochastic effect of quantum fluctuations during inflation can also introduce effective masses related to $H_\ast$ \cite{Linde:1985gh,Dolgov:1991fr,Dine:2003ax}.
If the scalar mass is associated with the Hubble scale of inflation, namely $m_\sigma \sim \mathcal{O}(H_\ast)$, we can find nearly constant masses in each phase where
\begin{align}\label{appro_constant_mass}
\nonumber
m_{1\pm}^2 &=  m_\sigma^2 -3 \frac{c_1 \pm c_2}{\Lambda} \sqrt{2\epsilon_{\rm CMB}} M_PH_\ast^2
+ \frac{c_3}{\Lambda^2} 2\epsilon_{\rm CMB} M_P^2H_\ast^2, \\
m_{2\pm}^2 &=  m_\sigma^2 + \frac{c_3}{\Lambda^2} 2\epsilon_{2} M_P^2H_\ast^2 \approx m_\sigma^2, \\\nonumber
m_{3\pm}^2 &=  m_\sigma^2 + \delta_2 \frac{c_1 \pm c_2}{\Lambda} \sqrt{2\epsilon_{\ast}} M_PH_\ast^2
+ \frac{c_3}{\Lambda^2} 2\epsilon_{\ast} M_P^2H_\ast^2,
\end{align}
where $\delta_3 = -\delta_2 - 3$ is used for $m_{3\pm}$.
The approximation $m_{2\pm} \approx m_\sigma^2$ is valid as long as $m_\sigma^2/H_\ast^2 \gg \epsilon_{\rm CMB}M_P^2/\Lambda^2$, which implies a lower bound for the cutoff $\Lambda \gg \epsilon_{\rm CMB}M_P$. 
In this case the time evolution of $\sigma_{\pm}$ is directly given by the results found in Section~\ref{Sec. massive_scalar_USR}.
Note that $m_\sigma \sim H_\ast$ also ensures that the density fluctuation of $\sigma$ during inflation merely contributes to negligible isocurvature perturbations in the CMB background radiations.

\bigbreak\noindent
\textbf{Massless AD field.}
Another usual assumption for the bare mass parameter $m_\sigma$ is of order the electroweak scale ($m_\sigma = m_{3/2}$), which has negligible contribution at the energy scale of inflation. The special massless limit ($m_\sigma \rightarrow 0$) can harvest a non-negligible baryon asymmetry at the end of inflation which is investigated in Appendix~\ref{Appd_massless_AD}.

\bigbreak
A possible scenario towards the UV completion of \eqref{Lagrangian_coupling} would be to consider a vector field $A^\mu$ that couples to the AD field and a scalar $\varphi$ through the interactions
\begin{align}\label{Lagrangian_UV}
\Delta\mathcal{L}_A \supset c_{A1} \varphi A^\mu \partial_\mu \varphi + c_{A2 } \sigma A^\mu \partial_\mu \sigma 
+ c_{A3} \sigma^\ast A^\mu \partial_\mu \sigma + c.c.,
\end{align}
where $\varphi$ should be identified with the inflaton via the redefinition $\Lambda_A \phi \equiv \varphi^2/2$. Since inflaton is real scalar, \eqref{Lagrangian_UV} cannot be a gauge theory. $A^\mu$ shall be heavy with a mass $m_A \sim \Lambda$ so that it can be integrated out at the energy scale of inflation. This results in the effective theory $c_{A1} \varphi A^\mu \partial_\mu \varphi \equiv c_{A1} \Lambda_A A^\mu \partial_\mu \phi 
\supset - c_{A1} c_{A2} \frac{\Lambda_A}{m_A^2} \sigma\partial^\mu\sigma \partial_\mu\phi -c_{A1}c_{A3} \frac{\Lambda_A}{m_A^2} \sigma^\ast\partial^\mu\sigma \partial_\mu\phi +\cdots$. The only choice for the effective theory to reproduce \eqref{Lagrangian_coupling} is to make sure $\Lambda_A = \Lambda$ where one can recognize $c_1 \sim c_{A1} c_{A3} $, $c_2 \sim c_{A1} c_{A2}$ and the $c_3$ term becomes higher-order corrections. Although there are analogous terms to $\frac{1}{\Lambda^2} \sigma^2 (\partial\sigma)^2$, $\frac{1}{\Lambda^2} \vert\sigma\vert^2 \vert\partial\sigma\vert^2$, $\cdots$ arise in this scenario that introduce non-canonical couplings to the mass eigenstates, one can check that these terms of $\Lambda^{-2}$ are indeed subdominant corrections to \eqref{Lagrangian_coupling} with the proper choice of the cutoff $H_\ast \ll \Lambda \leq M_P$.

There is a simple constraint $\mathcal{L}_\phi \approx 3M_P^2 H_\ast^2 \gg \Delta\mathcal{L}$ to ensure the de Sitter background of single-field inflation. The background assumption $\dot{\phi} \sim \sqrt{\epsilon_H}M_PH_\ast$ and the initial condition $\sigma_\pm(t_0) \sim H_\ast^2/m_\pm$ indicate that the $c_1$ or $c_2$ term has $\Delta\mathcal{L} \sim \sqrt{\epsilon_H} \frac{M_P}{\Lambda}\frac{H_\ast^6}{m_\pm^2}$, and the $c_3$ term has $\Delta\mathcal{L} \sim \epsilon_H \frac{M_P^2}{\Lambda^2}\frac{H_\ast^6}{m_\pm^2}$. Thus $\Delta\mathcal{L} \ll H_\ast^4 \ll \mathcal{L}_\phi$ is naturally consistent with the constant mass approximation $m_\sigma^2/H_\ast^2 \gg \epsilon_{\rm CMB}M_P^2/\Lambda^2$ used in \eqref{appro_constant_mass}.  

\subsection{Baryon number production}
The sudden transition of scalar masses induced by the change of inflaton rolling rate $\delta$ from slow-roll to USR triggers the coherent motion of the $\sigma$ VEV. As $\sigma$ starts in motion, the conserved current $j^\mu = i(\sigma^\ast \partial_\mu\sigma - \sigma\partial_\mu\sigma^\ast)$ associated with a global $U(1)$ quantum number in the limit of $\Delta\mathcal{L} \rightarrow 0$ is produced from zero. We will assign such a $U(1)$ to the baryon number for a typical example, despite that in general the $U(1)$ can be a combination of baryon, lepton or even quantum numbers in the dark sector \cite{Petraki:2013wwa,Bell:2011tn,vonHarling:2012yn}. In terms of mass eigenstates, the baryon number is
\begin{align}
n_B = j^0 = \sigma_{+}\dot{\sigma}_- - \sigma_{-}\dot{\sigma}_+ 
= H_\ast \left(\sigma_{+}\frac{d \sigma_{-}}{d N} - \sigma_{-}\frac{d \sigma_{+}}{d N}\right).
\end{align}
Note that it is only the $c_2$ term in \eqref{Lagrangian_coupling} that violates the baryon number.

To estimate the baryon asymmetry at the end of inflation, in this section we assume instantaneous reheating soon after $N_{\rm  end}$, where the radiation energy density $\rho_r(t_{\rm end}) \approx 3M_P^2 H_\ast^2$. The temperature $T(t) = [\frac{30}{\pi^2 g_\ast}\rho_r(t)]^{1/4}$ at the beginning of radiation domination is therefore given by $T_\ast \equiv T(H_\ast)$ and the corresponding entropy production $s_\ast = 2\pi^2g_\ast T_\ast^3/45$, where the number of relativistic degrees of freedom above $300$ GeV is $g_\ast = 106.75$.

Based on the results given by \eqref{late_time_Phase3} for each mass eigenstate, we can compute $n_B$ to find out the baryon asymmetry as
\begin{align}\label{Y_B__analyitc_inflation_end}
Y_B = \frac{n_B(N_{\rm end})}{s_\ast} =\frac{H_\ast}{s_\ast} D_+D_- 
\left(\Delta_{3+}^- - \Delta_{3-}^-\right) e^{-(\Delta_{3+}^- + \Delta_{3-}^-)N_{\rm end}},
\end{align}
where $D_\pm$ and $\Delta_{3\pm}^-$ are the coefficient \eqref{Phase3_coefficient} and the Phase 3 scaling dimension \eqref{Phase3_Delta} with respect to the mass eigenstates $m_{3\pm}$.
For a large duration number $N_{\rm end} \gg 1/(\Delta_{3+}^- + \Delta_{3-}^-)$, the exponential suppression of $Y_B$ illustrates the conventional expectation for a negligible baryon asymmetry from long-term inflation.

\begin{figure}
	\begin{center}
		\includegraphics[width=7cm]{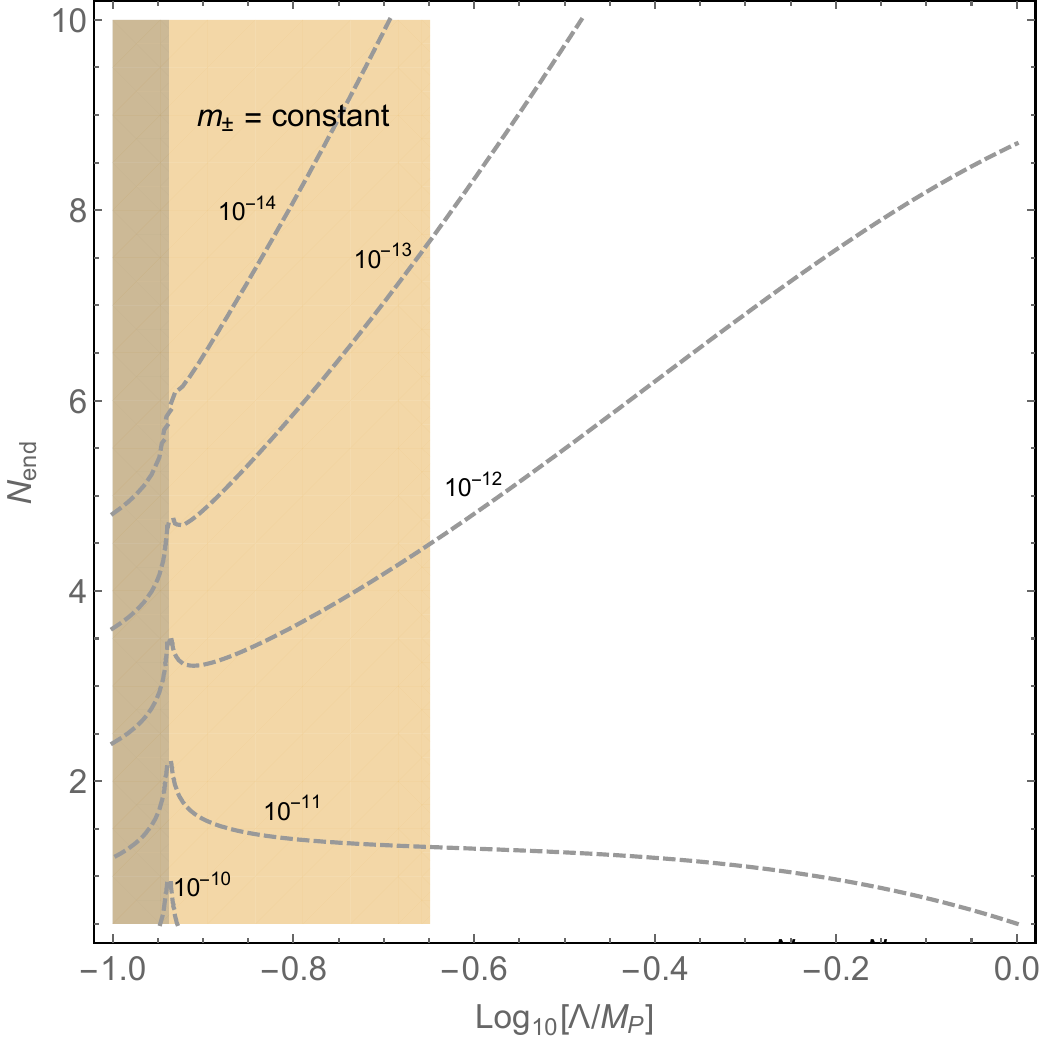} 
		%\par
	\end{center}
	\caption{The parameter scan for the baryon asymmetry $\vert Y_B \vert$ at the end of inflation from a heavy AD field ($m_\sigma = H_\ast/2$) in the USR limit $\delta_2 = -3$ with $\{c_1, c_2, c_3 \} = \{-2, 1, 1 \}$, $N_\ast = 0.5$ and $H_\ast = 2.37 \times 10^{13}$ GeV. Dashed lines are contours of $\vert Y_B\vert$. Colored regions are excluded by validity of the late-time approximation for $m_{3\pm}/H_\ast < 3/2$ and the constant-mass formalism \eqref{appro_constant_mass}. \label{fig:Y_B_heavy_AD}}
\end{figure}

%\subsubsection{Heavy AD field.}
Let us find out the typical $\vert Y_B \vert$ at the end of inflation from a heavy AD field.
For $m_\sigma \sim H_\ast$, we apply constant-mass approximation \eqref{appro_constant_mass} to compute the evolution of each mass eigenstate, where the solutions for $\sigma_{\pm}$ at the end of inflation are readily given in Section~\ref{Sec. massive_scalar_USR}.
With $H_\ast \sim 10^{13}$ GeV, the instantaneous reheating gives $H_\ast^3/s_\ast \sim 10^{-9}$ and with $\Lambda \sim M_P$, $D_+D_-/H_\ast^2 \sim  10^{-2}$ for arbitrary choices of $N_\ast$ and $N_{\rm end}$. $\vert \Delta_{3+}^- - \Delta_{3-}^-\vert \gtrsim 10^{-1}$ if $N_\ast \lesssim 0.5$ and $\vert \Delta_{3+}^- - \Delta_{3-}^-\vert \lesssim 10^{-3}$ if $N_\ast \gtrsim 2.5$. (Note that $N_\ast > 2.5$ is usually required for the byproduct PBHs from the USR inflation to be important dark matter, see the discussion Section~\ref{Sec.B_PBH_correlation}.) 

As a result, for heavy AD mass $0.2 < m_\sigma/H_\ast < 1.5$, our numerical results show that the maximal baryon asymmetry given by \eqref{Y_B__analyitc_inflation_end} is realized at the order of
\begin{align}
\left\vert Y_B \right\vert (m_\sigma, N_\ast, N_{\rm end} )\lesssim 10^{-11},
\end{align}
for short term USR inflation with $N_\ast < 0.5$ and $N_{\rm end} < 5$ that gives $e^{-(\Delta_{3+}^- + \Delta_{3-}^-)N_{\rm end}} \sim \mathcal{O}(1)$. 
A parameter scan for $\vert Y_B \vert$ at the end of inflation from a heavy AD field with $m_\sigma = H_\ast/2$ is shown in Figure~\ref{fig:Y_B_heavy_AD}. We conclude that for a heavy charged scalar with a bare mass $m_\sigma\sim H_\ast$, the observed baryon asymmetry cannot be purely explained by its dynamical excitation triggered by the USR phase transition during inflation. However, when taking into account the coherent oscillation of inflaton after the end of inflation in the usual (non-instantaneous) reheating scenario, a heavy scalar can still generate sufficient baryon asymmetry through the standard AD mechanism. We will explore  baryogenesis after inflation via reheating in Section~\ref{Sec. baryogenesis_after_inflation}.

\bigbreak\noindent
\textbf{The correlation length.}
Here we address the correlation length (or the variance) of the local baryon asymmetry associated with the initial conditions. First of all, it is remarkable that the choice of \eqref{initial_condition_Phase1} for the mass eigenstates in Phase 1 correspond to VEVs with maximal CP violation. The averaged baryon asymmetry among the probability distribution of all VEVs is expected to be dominated by those values with maximal CP violation  \cite{Wu:2020pej}. To see what is the meaning of maximal CP violation, let us apply the polar representation $\sigma = Re^{i\theta}/\sqrt{2}$ to the AD field, where the radial VEV is given by $\langle R^2 \rangle = \langle \sigma_{-}^2\rangle + \langle \sigma_{+}^2\rangle$ and the angular VEV can be computed via the relation $\tan \theta = \sigma_{+}/\sigma_{-}$. For heavy AD field considered here, $m_{1\pm}$ given by \eqref{appro_constant_mass} is dominated by $m_\sigma$ so that $\vert\sigma_{+}/\sigma_{-}\vert =m_{1-}/m_{1+} \approx 1$, which implies that the initial VEV of $\theta$ is approximately at $\pm \pi/4$ or $\pm 3\pi/4$. 

In terms of the polar representation, the equation of motion reads
\begin{align}\label{eom:R}
\ddot{R} + 3H\dot{R} +\left[m_\sigma^2 + \frac{c_1}{\Lambda}\square\phi + \frac{c_3}{\Lambda^2}(\partial\phi)^2 -\dot{\theta}^2  \right] R
=  \frac{c_2}{\Lambda}\square\phi R \cos(2\theta), \\
\label{eom:theta}
\ddot{\theta} + \left(3H +2 \frac{\dot{R}}{R}\right) \dot{\theta} = - \frac{c_2}{\Lambda}\square\phi \sin(2\theta).
\end{align}
%where the effective mass for the radial mode, $R$, is  $m_R(\theta)^2 = m_\sigma^2 + (c_1 - c_2 \cos (2\theta))\square\phi /\Lambda + c_3(\partial\phi)^2/\Lambda^2$. 
One can see that initial conditions with $\theta = \theta_{\rm max} \equiv \pm(2n+1) \pi/4 $ are the values that provide the maximal source for the $\theta$-motion on the right-hand-side of \eqref{eom:theta}. Given that $n_B = R \dot{\theta}$ and that the radial VEV is always positive, a local Universe filled with matter or antimatter is in fact determined by the initial VEV of $\theta$, which controls the sign of $\dot{\theta}$. 

We can derive the effective potential $V(R,\theta) =  m_R(\theta)^2 R^2/2$ for the AD field from \eqref{eom:R}, where $m_R(\theta)^2 = m_\sigma^2 + (c_1 - c_2 \cos (2\theta))\square\phi /\Lambda + c_3(\partial\phi)^2/\Lambda^2$. For a heavy AD field with $m_R \sim \mathcal{O}(H_\ast)$, the perturbation of the radial mode on top of the coherent value $R_0 \equiv \sqrt{\langle R^2\rangle}$ is negligible. In contrast to the presence of ``flat directions'' in higher-order non-renormalizable potential \cite{Dine:1995kz}, $\theta$ may not have a well-defined coherent value in this scenario, yet one can easily check that the effective mass of the angular mode at the maximal CP violation is given by $ m_\theta^2\sim \partial_{\theta}\partial_{\theta} V(R,\theta)\vert_{\theta = \theta_{\rm max}} \rightarrow 0$. 
The correlation length $x_c$ for a massive scalar in de Sitter, defined from $G(x_c) = G(0)/2$, is $x_c/x_{\rm ref} = 2^{3H_\ast^2/(2m^2)}$ \cite{Wu:2020ilx,Starobinsky:1994bd}, where $m$ is the scalar mass, $x_{\rm ref}$ is a reference length scale and $G(x)\equiv G(\vert \vec{x}_1 -\vec{x}_2\vert)$ is the two-point spatial correlation function.
Since we are interested in long wavelength modes that have exited the horizon by the time of USR transition 
(namely $x_{\rm ref} \sim 1/k_{\rm CMB}$ can be a good choice), 
the condition $m_\theta \rightarrow 0$ implies that baryon asymmetry from $\theta$ picked up by the choice \eqref{initial_condition_Phase1} for the mass eigenstates has a correlation length much larger than the Hubble scale, which ensures that a local patch of the Universe is left with a pure (anti)matter (see also the discussion in Section 4.1 of Ref. \cite{Dine:1995kz}). 

\section{Baryogenesis after inflation}\label{Sec. baryogenesis_after_inflation}
The presence of a coherent oscillation of the inflaton following the end of the (secondary) slow-roll phase is a conventional assumption to terminate inflation. Once the inflaton $\phi$ rolls into a deep potential valley, it acquires a large mass $m_I \gg H_\ast$ and starts to oscillate around the potential minimum, leading to an effective matter-domination epoch. Reheating starts with the decay of inflation into radiative degrees of freedom and finally the energy density of radiation dominates the Universe. In this section, we investigate the subsequent evolution of the AD field from inflaton coherent oscillations to completed reheating  (or namely the beginning of radiation domination). 

\begin{figure}[]
	\begin{center}
		\includegraphics[width=9 cm]{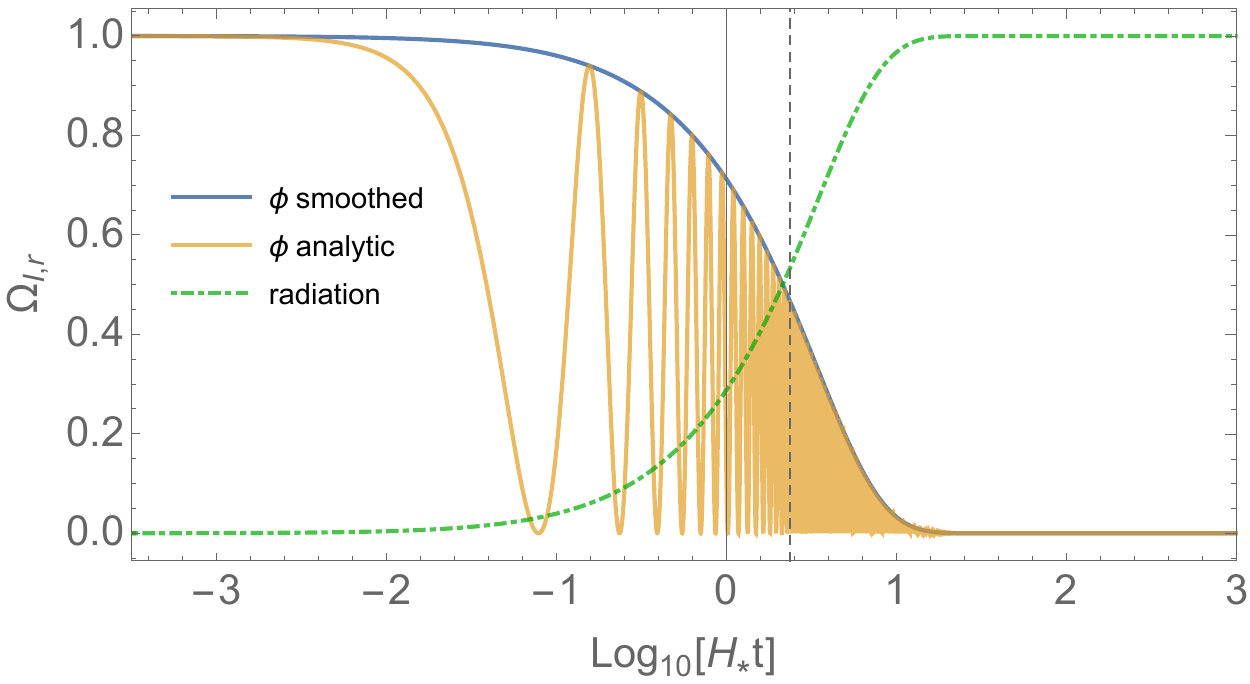}
	\end{center}
	\caption{\label{fig.inflaton_density} The inflaton energy density $\Omega_I \equiv \rho_I/(3M_p^2 H^2)$ as a function of time based on the smoothed and analytic approximations with an inflaton mass $m_I = 20 H_\ast$. The dotted-dashed line depicts the energy density of radiation $\Omega_r \equiv \rho_r/(3M_p^2 H^2)$. The vertical dashed line is $t = 1/\Gamma_I$.}
\end{figure}

We assume the existence of a deep valley $\partial^2 V(\phi)/\partial\phi^2 \approx m_I^2 \gg H_\ast^2$ without specifying the shape of the potential $V(\phi)$. The decay of $\phi$ is governed by a perturbative channel, featured by the decay width $\Gamma_I$ of the inflaton, where the energy density of inflaton and radiation ($\rho_I$ and $\rho_r$) evolve as
\begin{align}
\dot{\rho}_I + 3H\rho_I = -\Gamma_I \rho_I, \quad \dot{\rho}_r +4H\rho_r = \Gamma_I \rho_I.
\end{align} 
The evolution of each density species satisfies the constraint $3M_P^2 H^2 = \rho_I + \rho_r$. It is convenient to defined the energy scale of inflation $\Lambda_I^4 = 3M_P^2H_\ast^2 = \rho_{I0}$ as the initial condition for the post-inflationary dynamics, where $\rho_{I0} = \rho_I(t_{\rm end})$.  The energy density of radiation is created from zero: $\rho_{r0} = 0$.

To mimic the background expansion in the effective matter domination Universe driven by the rapid oscillation of $\phi$, we impose a simplified analytic representation \cite{Wu:2019ohx} as
\begin{align}\label{analytic_inflaton}
 \phi(t) = \phi_{\rm max} a^{-3/2} \cos\left[m_I (t-t_{\rm end})\right] e^{-\Gamma_I (t-t_{\rm end})/2}, 
\end{align}
where $\phi_{\rm max} \equiv \Lambda_I^2/m_I$ is the maximal oscillatory amplitude for $\phi$ at the beginning of oscillation. This analytic representation is a good approximation as long as $m_I \gg H_\ast$ and $H_\ast(t-t_{\rm end}) \gg 1$. For clearness and simplicity we take $t_{\rm end} = 0$ from now on. In the limit of $m_I \rightarrow \infty$, one can reproduce the smoothed energy density of the inflaton as $\rho_I = \langle m_I^2 \phi^2 \rangle \rightarrow \Lambda_I^4 a^{-3} e^{-\Gamma_I t}$. We plot in Figure~\ref{fig.inflaton_density} an analytic example with $m_I = 20 H_\ast$. One can see that $t_r \equiv 1/\Gamma_I$ is the approximated time scale at the beginning of radiation domination for arbitrary choice of $m_I$.

Based on the analytic representation \eqref{analytic_inflaton} we can express the mass terms \eqref{mass_pm} of the mass eigenstates $\sigma_{\pm}$ as explicit functions of the background dynamics through $\square\phi = -\ddot{\phi} -3H\dot{\phi} = m_I^2\phi$ and $(\partial\phi)^2 = \dot{\phi}^2$. Therefore we are now ready to study the time evolution of $\sigma_{\pm}$ in the post-inflationary epochs with initial conditions at the end of inflation investigated in the previous sections.\footnote{The AD field has a maximal energy density at the end of inflation where the $c_1$ or $c_2$ term has $\Delta\mathcal{L} \lesssim \frac{1}{\Lambda} \frac{H_\ast^4}{m_\sigma^2}m_I^2\phi_{\rm max} \sim \frac{m_I \Lambda_I^2}{\Lambda m_\sigma^2} H_\ast^4$ and the $c_3$ term has $\Delta\mathcal{L} \lesssim \frac{\Lambda_I^4}{\Lambda^2} \frac{H_\ast^4}{m_\sigma^2}$ as $\sigma_{\pm} \sim H_\ast^2/m_\sigma$ is slightly decay during inflation. For a heavy AD field with $m_\sigma \sim H_\ast$, $\Delta\mathcal{L} \ll \mathcal{L}_\phi$ holds in an apparent way so that reheating driven by the coherent oscillation of $\phi$ is not interrupted by the AD field back reactions.
}

%\subsection{Dynamical initial conditions}\label{Sec: dyn_I}
\bigbreak
\noindent
\textbf{Dynamical initial conditions.}
As seen from \eqref{late_time_Phase3}, for the ending time of inflation $N_{\rm end} \lesssim 1/\Delta_{3\pm}^- $, VEVs of the AD field excited by the USR phase transition remains in motion, providing dynamical initial conditions for their relaxation during reheating.
% In this section we investigate the resulting baryon asymmetry from this class of initial conditions.

\begin{figure}
	\begin{center}
		\includegraphics[width=6.5cm]{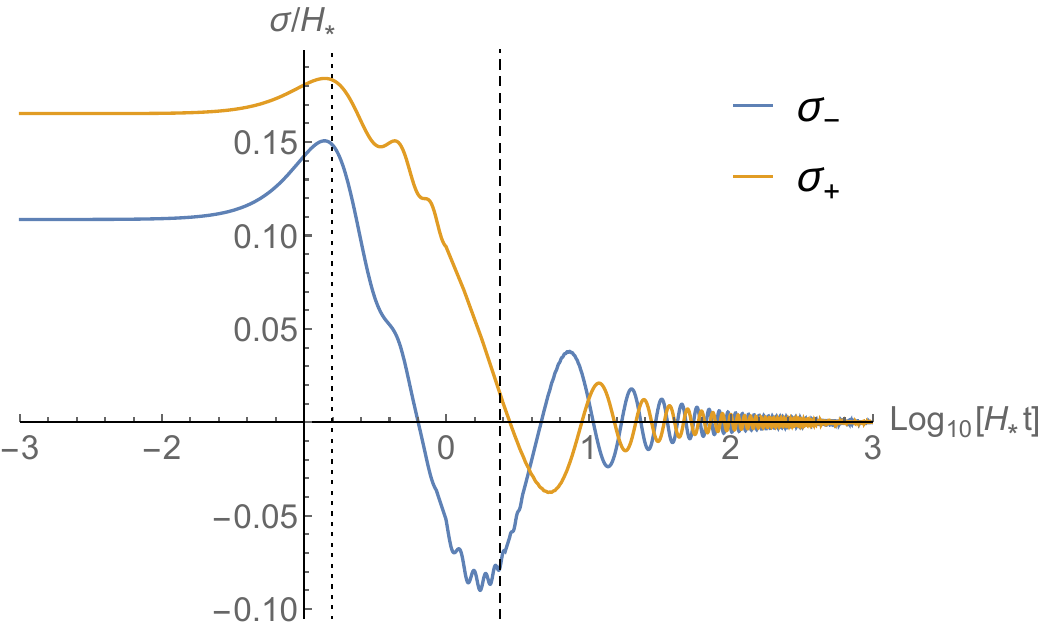} 
		\hfill
		\includegraphics[width=6.5cm]{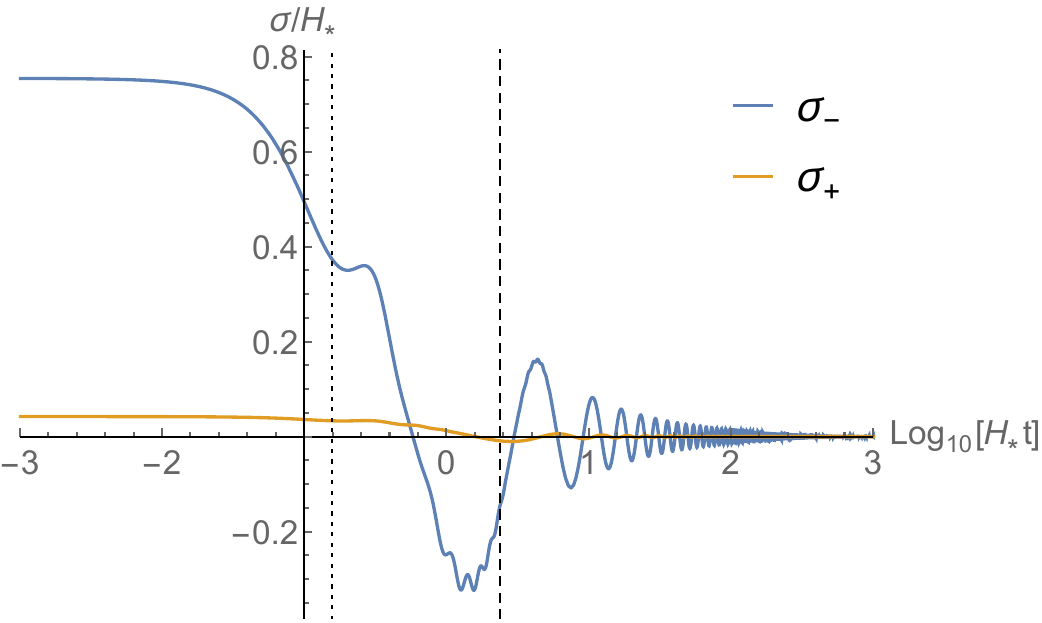} 
		%\par
	\end{center}
	\caption{The time evolution of mass eigenstates $\sigma_{\pm}$ from the end of inflation to radiation domination with $m_I = 20 H_\ast$, $\Lambda = M_P$, $H_\ast = 2.37 \times 10^{13}$ GeV and $\Gamma_I = 10^{13}$ GeV. In both panels $N_\ast = 1$ and $N_{\rm end} = 5$ are used. The  vertical dotted lines are $t = \pi/m_I$ and the vertical dashed lines are $t = 1/\Gamma_I$.
		Left Panel: The case of tachyonic initial masses with $\{c_1, c_2, c_3 \} = \{-2, 1, 1 \}$ and $m_\sigma = H_\ast/2$.    
		Right Panel: The case of positive masses with $\{c_1, c_2, c_3 \} = \{2, -1, 1 \}$ and $m_\sigma = H_\ast$. 
		\label{fig:masseigen_dynamic}}
\end{figure}

Let us focus on a heavy AD field with $m_\sigma \sim H_\ast$ and $\Lambda \sim M_P$ which guarantees initial conditions of $\sigma_{\pm}$ that are well computed by the constant-mass approximation \eqref{appro_constant_mass}. For $m_\sigma \lesssim H_\ast$ and $m_I \gg H_\ast$, the initial masses at $t_{\rm end}$ are dominated by $c_1$ and $c_2$ terms where $m_\pm^2 \approx (c_1\pm c_2) m_I \Lambda_I^2/\Lambda \sim (c_1\pm c_2) m_I H_\ast \gg H_\ast^2$. This is due to the transition of potential energy to kinetic terms for the inflaton at the end of inflation, which gives the opportunity for $m_\pm$ to become tachyonic at the beginning of reheating, depending on the choices of $c_1$ and $c_2$. 
In Figure~\ref{fig:masseigen_dynamic}, we show examples for the evolution of $\sigma_{\pm}$ from tachyonic initial masses $m_\pm^2(t_{\rm end}) < 0$ and positive initial masses $m_\pm^2(t_{\rm end}) > 0$ with different choices of the coefficients $c_i$. For both choices of $c_1$ and $c_2$ in Figure~\ref{fig:masseigen_dynamic}, the effective masses $m_{i \pm}^2$ during inflation given by \eqref{appro_constant_mass} are all positively defined. It is remarkable that these initial VEVs are in the regime of $\sigma_{\pm} \ll H_\ast$, while the conventional AD baryogenesis from flat directions \cite{Dine:1995kz} relies on initial VEVs much greater than the scale of $H_\ast$.

\begin{figure}
	\begin{center}
		\includegraphics[width=7cm]{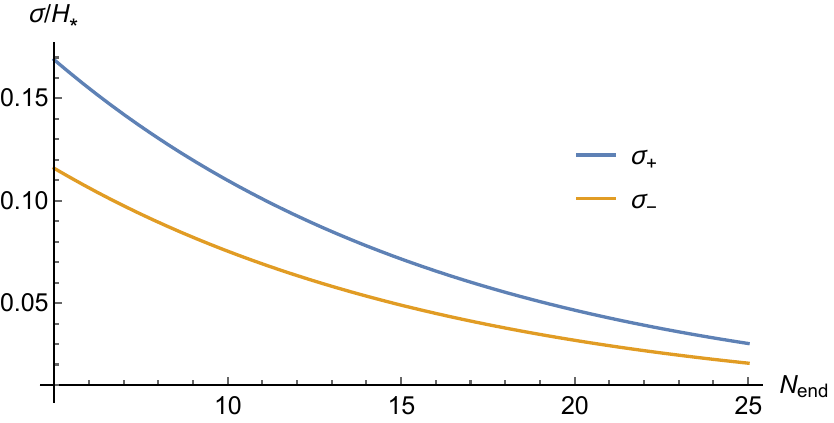} 
		\hfill
		\includegraphics[width=7cm]{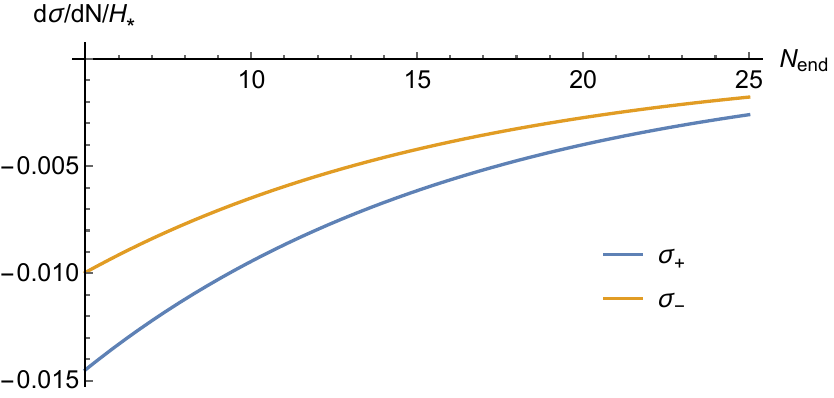} \\\bigbreak
		\includegraphics[width=8cm]{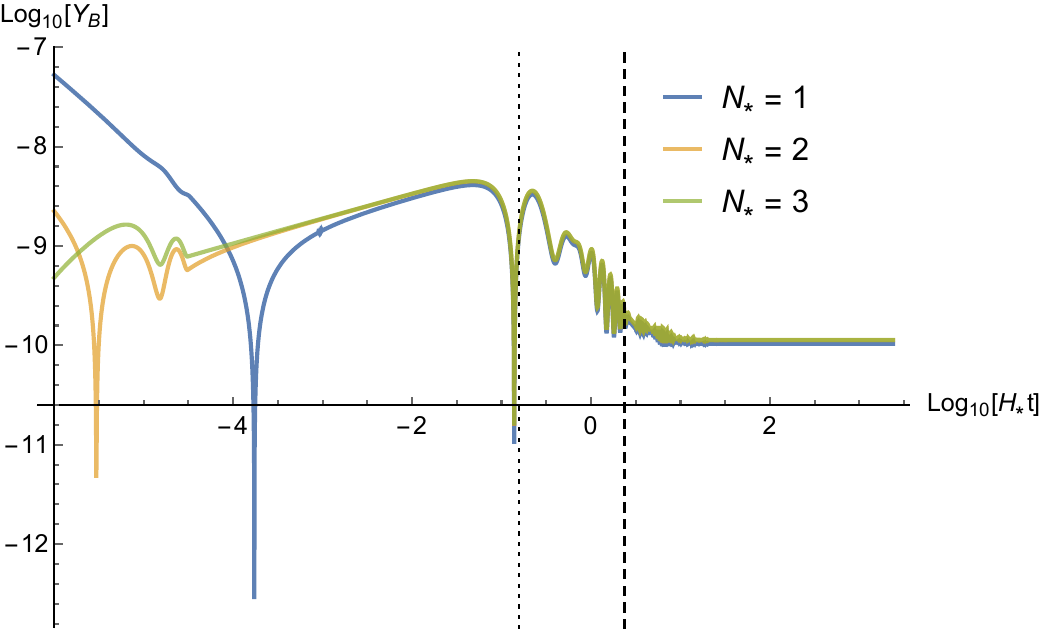} 
		\hfill
		\includegraphics[width=6.5cm]{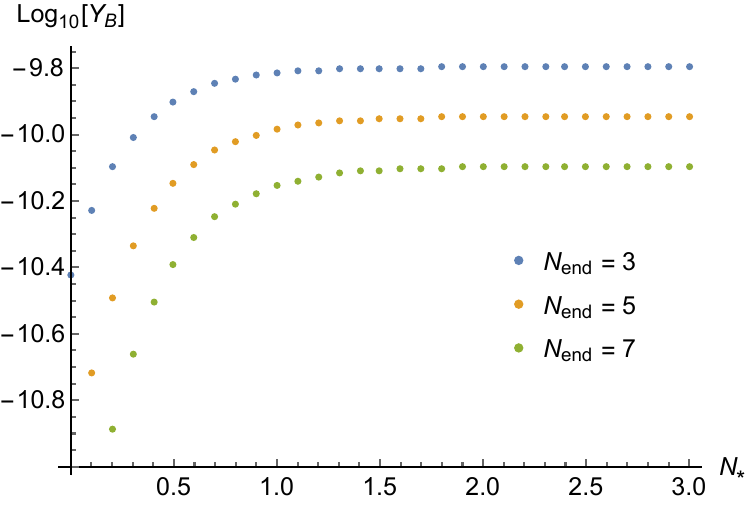} 
		%\par
	\end{center}
	\caption{The baryon asymmetry $\vert Y_B \vert$ generated by a heavy AD field $m_\sigma = H_\ast/2$ with tachyonic initial masses led by the choice of $\{c_1, c_2, c_3 \} = \{-2, 1, 1 \}$. 
		%Initial conditions for $\sigma_{\pm}$ are picked up in the dynamical region $\Delta_{3\pm}^- N_{\rm end} \sim \mathcal{O}(1)$. 
		Upper Panel: Initial conditions of the mass eigenstates (left) and their time derivatives (right) with respect to $N_{\rm end}$.
		Lower Left Panel: The time evolution of $\vert Y_B \vert$ from the end of inflation to reheating complete with respect to various choices of $N_\ast$ and $N_{\rm end} = 5$. The  vertical dotted line is $t = \pi /m_I$ and the vertical dashed line is $t = 1/\Gamma_I$. Lower Right Panel: The final $\vert Y_B \vert$ in radiation domination with respect to various choices of $N_{\rm end} $. In all panels, $\Lambda = M_P$, $H_\ast = 2.37 \times 10^{13}$ GeV and $\Gamma_I = 10^{13}$ GeV are used. \label{fig:Y_B_dynamics_Ns}}
\end{figure}

%\begin{figure}
%	\begin{center}
%		\includegraphics[width=8cm]{Y_B_dyn_positive_Ns.pdf} 
%		\hfill
%		\includegraphics[width=7cm]{Y_B_dyn_positive_Nend.pdf} 
		%\par
%	\end{center}
%	\caption{The baryon asymmetry $\vert Y_B \vert$ generated by a heavy AD field $m_\sigma = H_\ast$ with positive initial masses led by the choice of $\{c_1, c_2, c_3 \} = \{2, -1, 1 \}$. Initial conditions for $\sigma_{\pm}$ are picked up in the dynamical region $\Delta_{3\pm}^- N_{\rm end} \sim \mathcal{O}(1)$. 
%		Left Panel: The time evolution of $\vert Y_B \vert$ from the end of inflation to reheating complete with respect to various choices of $N_\ast$ and $N_{\rm end} = 5$. The  vertical dotted line is $t = \pi /m_I$ and the vertical dashed line is $t = 1/\Gamma_I$. Right Panel: The final $\vert Y_B \vert$ in radiation domination with respect to various choices of $N_{\rm end} $. In both panels, $\Lambda = M_P$, $H_\ast = 2.37 \times 10^{13}$ GeV and $\Gamma_I = 10^{13}$ GeV are used. \label{fig:Y_B_dyn_positive}}
%\end{figure}

With the numerical solutions of the mass eigenstates $\sigma_{\pm}$, we can obtain the baryon asymmetry $Y_B(t) = n_B(t)/s(t)$ during reheating where $n_B = \sigma_{+}\dot{\sigma}_- - \sigma_{-}\dot{\sigma}_+$.
%The resulting baryon asymmetry $Y_B$ from the tachyonic initial masses of $m_\pm$ at the end of inflation are given in Figures~\ref{fig:Y_B_dynamics_Ns} and \ref{fig:Y_B_dyn_positive}. 
For arbitrary choices of the initial conditions, the ending time of inflation, $N_{\rm end}$, is the most important parameter that determines the final $Y_B$ in radiation domination. Some examples based on the tachyonic initial masses of $m_\pm$ at the end of inflation are given in Figure~\ref{fig:Y_B_dynamics_Ns}.
 The decay of $\vert Y_B \vert$ due to the increase of $N_{\rm end}$ in the initial conditions is easy to understand since the initial VEVs of $\sigma_{\pm}$ and their time derivative are exponentially diluted by $N_{\rm end}$. 

On the other hand, $Y_B$ is not sensitive to the duration of the USR phase, $N_\ast$, from arbitrary initial conditions. This is due to the fact that with sufficiently large $N_\ast$, $\epsilon_{\ast} = \epsilon_{\rm CMB} e^{2\delta_2 N_\ast}$ becomes too small so that the effective masses $m_{2\pm} = m_{3\pm} \approx m_\sigma$, as seen from \eqref{appro_constant_mass}. As a result, initial conditions of the mass eigenstates at the end of inflation are nearly the same for $N_\ast \gtrsim 1$ when $\delta_2 = -3.05$. Such an asymptotic constant behavior for the final baryon asymmetry towards the large $N_\ast$ limit has an important implication to the PBH formation from the USR inflation models. As will be discussed in Section \ref{Sec.B_PBH_correlation}, $N_\ast > 2$ is generally required for the by product PBHs from USR inflation to act as important dark matter today.

%Parametric resonance of the $\sigma_{\pm}$ amplitude can happen in certain parameter space due to the coherent oscillation of $\phi$. Without the appearance of parametric resonance, $\sigma_{\pm} \sim a^{-3/2}$ decay as massive scalars and the final expectation value of $Y_B$ in general satisfies the relation   
%\begin{align}
%\left\vert Y_B \right\vert \sim \frac{n_B(t_c)}{s(t_r)} \left(\frac{a_c}{a_r}\right)^3,
%\end{align}
%where $t_c \equiv \pi/m_I$ is the first period at which $m_\pm \approx (c_1\pm c_2)m_I^2\phi$ oscillate across zero and change their signs. $t_r \sim 6/\Gamma_I$ is the approximated epoch at the beginning of a fully radiation domination. Due to the growth (or decay) of $\sigma_{\pm}$ from the tachyonic (or positive) initial masses of $m_\pm$ given in Figure~\ref{fig:masseigen_dynamic}, the value of $n_B(t_c)$ is only determined numerically.

The decay of AD condensate due to scattering with thermalized particles during reheating could be an important issue, especially for baryogenesis following high-scale inflation considered in this work. It is usually assumed that the finite temperature effect induced a thermal mass $m_T$ to the AD field led by the smallest Yukawa coupling $y_s$ as $m_T \sim y_s T$ \cite{Anisimov:2000wx}. For the scale of inflation $\Lambda_I \sim 10^{16}$ GeV used in our examples, the maximal temperature during reheating is $T_{\rm max} \sim (M_P \Lambda_I^2 \Gamma_I)^{1/4} \sim 10^{15}$ GeV. However, one can check that if $y_s$ is of order $10^{-2} - 10^{-4}$, $m_T$ has negligible contribution to the relaxation process since $m_\pm \lesssim (m_I \Lambda_I^2/\Lambda)^{1/2}$ is dominated by inflaton kinetic energy at the beginning of reheating. Therefore, unless $y_s$ could be of order $10^{-1}$ or larger, we find that the finite temperature effect does not play an important role in the AD mechanism.

%\subsection{Long-term inflation}
\bigbreak
\noindent
\textbf{Long-term inflation.}
For models of inflation that experience a long duration until the end of the secondary slow-roll phase with $N_{\rm end} \gg 1/\Delta_{3\pm}^- $, the coherent motion of the AD field triggered by the Phase 1 to 2 transition becomes negligibly small. In this case the accumulation of long wavelength perturbations that exit the horizon in Phase 3 can develop new condensates for the mass eigenstates $\sigma_{\pm}$. As the subsequent scalar condensate reaches equilibrium under the stochastic effect \cite{Starobinsky:1994bd}, the VEV of $\sigma_{\pm}$ can be estimated by
\begin{align}\label{VEV_long_term}
\sigma_{\rm end \pm} = \sqrt{\frac{3}{8\pi}} \frac{H_\ast^2}{m_{\rm end \pm}}, \qquad
\dot{\sigma}_{\rm end \pm} = 0,
\end{align}
where $\epsilon_{\rm end} \sim \mathcal{O}(1)$, and for convenience we take $\epsilon_{\rm end} = 1/2$ so that \eqref{mass_pm} at $N_{\rm end}$ gives
\begin{align}
m_{\rm end\pm}^2 =  m_{3 \pm}^2(t_{\rm end}) = m_\sigma^2 + \frac{c_1 \pm c_2}{\Lambda}\delta_2 M_PH_\ast^2
+ \frac{c_3}{\Lambda^2}  M_P^2H_\ast^2.
\end{align}
Note that $m_{\rm end\pm}$ is a constant in constant-rate inflation with $\delta_2 \leq -3$.

\begin{figure}
	\begin{center}
		%\includegraphics[width=7cm]{Y_B_cutoff_longterm.pdf} 
		%\hfill
		\includegraphics[width=8cm]{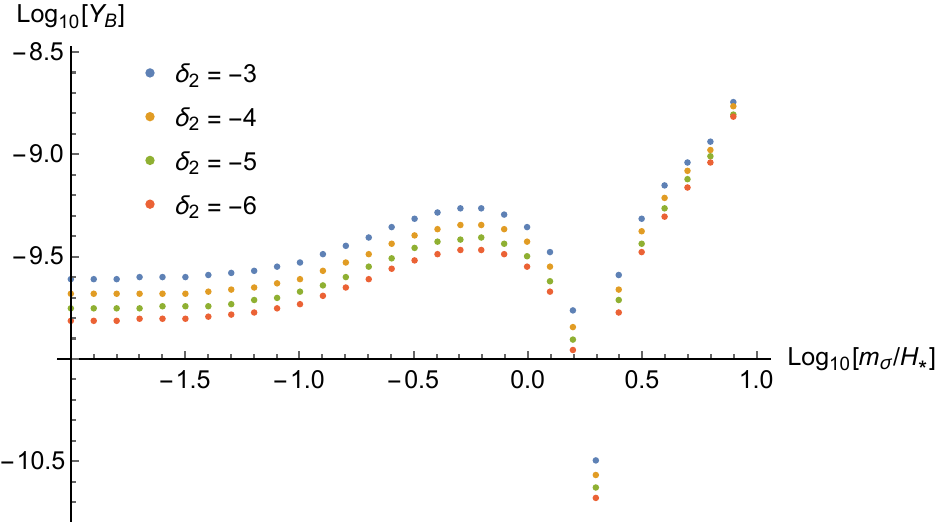} 
		%\par
	\end{center}
	\caption{The baryon asymmetry $\vert Y_B \vert$ from static initial conditions \eqref{VEV_long_term} after long-term inflation with respect to the mass parameter $m_\sigma$ of the AD field and various choices of $\delta_2$. In this plot, we use $\{c_1, c_2, c_3 \} = \{-2, 1, 1 \}$, $m_I = 20H_\ast$, $\Lambda = 0.3 M_P$, where $H_\ast = 2.37 \times 10^{13}$ GeV and $\Gamma_I = 10^{13}$ GeV. \label{fig:Y_B_longterm}}
\end{figure}

Baryogenesis from the static VEVs given by \eqref{VEV_long_term} does not rely on the constant-mass approximation \eqref{appro_constant_mass} nor the late-time expression \eqref{late_time_Phase3}, so that the applicable parameter space for $m_\sigma$ and the cutoff scale $\Lambda$ is less constrained. This allows us to explore  cases away from the USR limit with $\delta_2 < -3$. In Figure~\ref{fig:Y_B_longterm} we compare the resulting $Y_B$ with different choices of $\delta_2$ away from the USR limit. The final $Y_B$ approaches to a constant value in the massless limit where $m_\sigma \ll m_{\rm end\pm}$. The enhancement of $Y_B$ in the large AD mass limit $m_\sigma \gg H_\ast$ is due to the resonance amplification led by the choice of inflaton mass $m_I/H_\ast \sim \mathcal{O}(10)$. Such a resonance between inflaton and the AD field could happen in reality but is not a necessary condition for the scenario to reach enough baryon asymmetry.
We remark that the purpose of this section is to show a complete analysis of the possible initial conditions for baryogenesis triggered by constant-rate inflation.
However, information related to parameters for PBH formation, such as $N_\ast$ or $N_{\rm end}$, will be washed away by the long-term inflation. 
For the discussion of the PBH-baryon correlation in Section~\ref{Sec.B_PBH_correlation}, we shall focus on dynamic initial conditions. 
%given in Section~\ref{Sec: dyn_I}.

\section{Baryon-PBH correlation}\label{Sec.B_PBH_correlation}
So far we have investigated baryon production during or after multi-stage USR inflation without taking into account  PBH formation is generated by the enhanced power spectrum $P_\zeta$ on scales much smaller than those of the CMB. As the inflaton decays during reheating, the enhanced small-scale curvature perturbation $\zeta$ is inherited by the density perturbations of the radiation, and high peaks of these superhorizon density perturbations collapse to form PBHs when they reenter the horizon. For those byproduct PBHs from USR inflation to occupy a significant fraction of the dark matter density $\Omega_{\rm DM}$ at matter-radiation equality ($t =t_{\rm eq}$), the power spectrum $P_\zeta$ has to satisfy several constraints set by current observations. In this section, we investigate the viable parameter space for the multi-stage USR inflation to produce significant PBH dark matter and a large enough baryon asymmetry.

\subsection{PBH dark matter}
We focus on the viable window for PBHs to be all dark matter in the asteroid-mass range $M_{\rm PBH}/M_\odot \simeq 10^{-16} - 10^{-12}$. This indicates that the pivot scale $k_0/k_{\rm eq} \approx$ $(M_{\rm eq}/M_{\rm PBH})^{1/2} $ for the initiation of the USR phase is subject to the range of $k_0 \simeq 10^{12} - 10^{14}$ Mpc$^{-1}$, where $k_{\rm eq} = 0.01$ Mpc$^{-1}$ and $M_{\rm eq} \simeq 2.94\times 10^{17} M_\odot$  
%and $g_{\rm eq} = 3$ 
are the horizon wavenumber and the horizon mass at matter-radiation equality, respectively. The comoving horizon length $R_H \equiv 1/(a H)$ as a function of the horizon mass $M_H = 4\pi \rho(H) H^{-3}/3$ is given by 
\begin{align}
R_H(M_H) = \frac{1}{k_{\rm eq}}\left(\frac{M_H}{M_{\rm eq}}\right)^{1/2}\left(\frac{g_\ast}{g_{\rm eq}}\right)^{1/6},
\end{align}
where $g_{\rm eq} = 3$ accounts for the relativistic degrees of freedom at $t_{\rm eq}$.  Note that $R_{\rm eq} = R_H(M_{\rm eq}) \approx 1.57\times 10^{40}$ GeV$^{-1}$.

We define $\Omega_{\rm PBH}(R_H) = \Omega_{\rm PBH}(M_H)$ as the density parameter of PBHs at a given scale $R_H$ (or namely $M_H$) in radiation domination. The mass fraction, $\beta(M_{\rm PBH}, M_H) \equiv d\Omega_{\rm PBH}/ d \ln M_{\rm PBH}$, describes the distribution of PBH masses $M_{\rm PBH}$ at the given scale $M_H$. The statistics of the PBH density at the comoving scale $R_H$ can be summarized as
\begin{align}\label{PBH_density_general}
\Omega_{\rm PBH}(R_H) = \int\cdots\int_{F_c}^{\infty} \frac{M_{\rm PBH}}{M_H} f_c(y_i) P(F, y_i, \sigma_{Fi}) dF dy_1\cdots dy_i,
\end{align}
where a Gaussian random field is an ideal choice for $F$ and $y_i$ denotes spatial or temporal derivatives (constraints) of $F$. 
$f_c(y_i)$ describes spacetime constraints to ensure that the selected $F$ meets the criteria of PBH formation \cite{Bardeen:1985tr,Young:2014ana,Green:2004wb,Suyama:2019npc,Wu:2020ilx}.
$\sigma_{Fi}$ describes the $i$-th spectral moment of $F$ which can be computed by the power spectrum of $F$ as
\begin{equation}\label{spectral_moment}
\sigma_{Fi}^2(R_H) = \int_{0}^{\infty} k^{2i} W^2(kR_H) P_{F}(k) d\ln k,
\end{equation} 
where $W(kR_H)$ is a window function smoothing over $R_H$ to prevent the divergence in the limit of $k\rightarrow \infty$.
 The probability distribution function $P(F, y_i, \sigma_{Fi})$ is in general a multi-variable structure due to the non-trivial conditional distribution of $F$. The density contrast of radiation is the most widely adopted choice of $F$ for PBH formation \cite{Carr:1975qj}. The curvature perturbation $\zeta$ is another natural option for the fundamental random field in the statistics of PBH abundance from models of inflation \cite{Yoo:2018esr,DeLuca:2019qsy,Kalaja:2019uju}. The compaction function is a good choice of $F$ to explore the $P(F, y_i, \sigma_{Fi})$ correlation with the threshold value $F_c$ \cite{Germani:2019zez,Young:2020xmk}, and the inflaton perturbation, $\delta\phi$, is a convenient candidate of $F$ to study the effect of quantum diffusion in ultra-slow-roll scenarios \cite{Biagetti:2021eep}. 

\subsubsection{The fiducial statistics}
The simplest one-variable Gaussian statistics, referred to as the Press-Schechter method \cite{Carr:1975qj}, is the standard approach to obtain a reference PBH abundance from models of inflation. As the fiducial method to forecast the PBH abundance from the constant-rate model \eqref{CR_template}, we adopt the Press-Schechter formula for the mass fraction of the form: 
\begin{align}\label{beta_Press_Schechter}
\beta_{\rm PS}(M_H) = 2 \int_{\nu_c}^{} \frac{e^{-\nu^2/2}}{\sqrt{2\pi}} d\nu = \textrm{erfc}\left(\nu_c/\sqrt{2} \right), 
\end{align}
where $F$ is identified as the density contrast of radiation, $\nu \equiv F/\sigma_{F0}$ is the peak value for a Gaussian random field $F$ over its variance $\sigma_{F0}$ (or the zeroth spectral moment), and $\nu_c \equiv F_c/\sigma_{F0}$ is the critical peak value defined at the threshold $F_c$ of gravitational collapse. The analytic expression of \eqref{beta_Press_Schechter} is given in Appendix~\ref{Appd:PS_method}.

Given that PBHs behave as dust-like matter, the mass fraction grows as $\beta(M_H) = \beta(M_{\rm eq}) (a/a_{\rm eq})$ in the radiation-dominated Universe. The final PBH density at matter-radiation equality is therefore given by
\begin{equation}\label{PBH_density_eq}
\Omega_{\rm PBH, eq} = \int \beta( M_H) \left(\frac{M_{\rm eq}}{M_H}\right)^{1/2} d\ln M_H
\equiv \Omega_{\rm DM,eq}\int f(M_H) d\ln M_H
\end{equation}  
where the scaling relation $M_H/M_{\rm eq}\sim a/a_{\rm eq}$ is used. The mass function $f(M_H) = \beta( M_H)$ $(M_{\rm eq}/M_H)^{1/2}/\Omega_{\rm DM,eq}$ shows the PBH ratio in dark matter density at each horizon scale $M_H$ and $f_{\rm PBH} = \Omega_{\rm PBH, eq}/ \Omega_{\rm DM,eq}$ measures the contribution from all PBHs in the dark matter density.

Let us explore the viable parameter space for realizing $f_{\rm PBH} > 0.1$, where PBHs occupy a significant fraction of the dark matter density. A perfect USR inflation with $\delta_2 = -3$ will launch a scale-invariant power spectrum from $k_0$ all the way to $k_{\rm end}$. This means that the dominant contribution to the mass function $f(M_H)$ at matter-radiation equality comes from the mass range around  $M_H \gtrsim 10^{15}$ gram, as lower mass PBHs have been fully evaporated via Hawking radiation. The perfect USR case with $f_{\rm PBH} > 0.1$ is thus immediately ruled out by the severe constraints on PBH evaporation in the CMB, and the extra and inner galactic background radiation.

\begin{figure}
%\hspace*{-.3in}
\centering
\includegraphics[width=0.7 \textwidth]{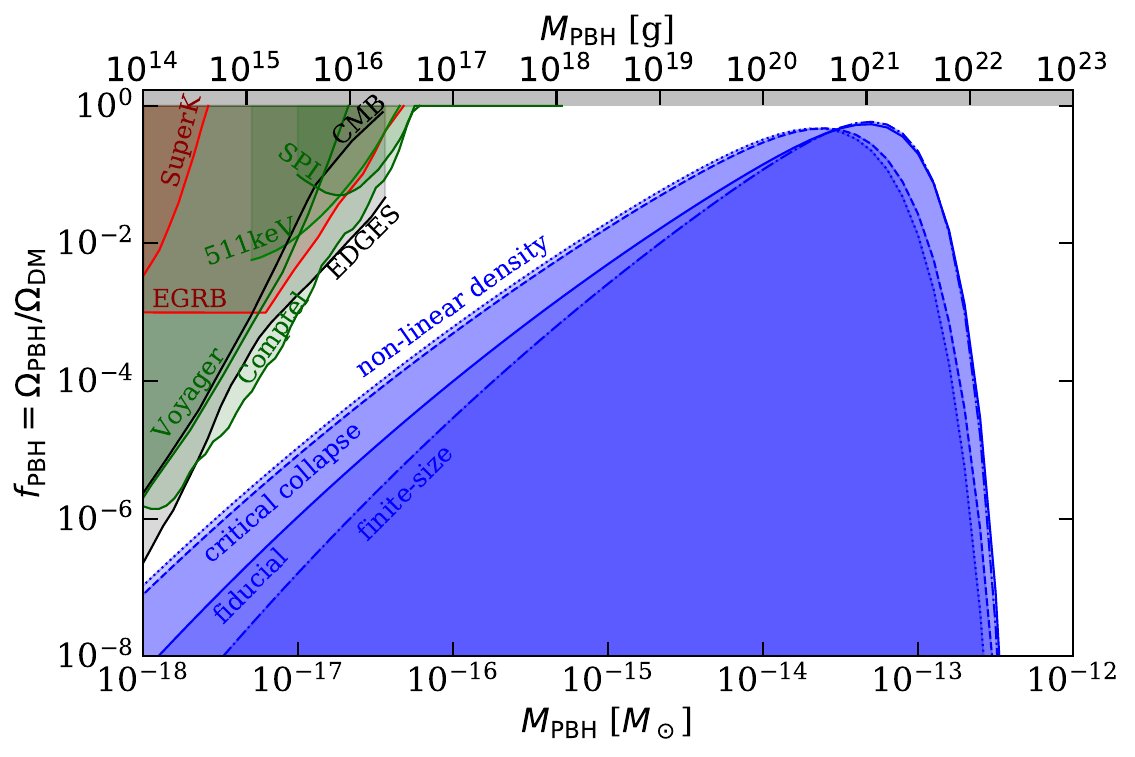}
\caption{
	The PBH mass function $f(M_{\rm PBH})$ (blue) based on the fiducial (Press-Schechter) approach (\ref{Appd:PS_method}, solid line) and extended mass functions given by the effect of critical collapse (\ref{Appd:critical_collapse}, dashed), the finite-size effect (\ref{Appd:finite_size}, dash-dotted) and non-linear density contrast (\ref{Appd:NL_density}, dotted). The pivot scale $k_0 = 7.3\times 10^{12}$ Mpc$^{-1}$ for the USR transition is used, and $N_\ast$ in each function is fixed by $f_{\rm PBH} =1$ with $\delta_2 = -3.05$ and $N_{\rm end} = 20$. The existing observational bounds are shown from: Galactic (red) and extragalactic (green) probes, as well as CMB and {\sc Edges} (black).}
	 \label{fig:f_k12}
\end{figure}

Fortunately, a slightly red-tilted USR inflation with $\delta_2 \lesssim -3$ can lead to sufficient decay of $f(M_{\rm PBH})$ towards the small mass limit. In Figure~\ref{fig:f_k12} we provide an example of the mass function computed by the fiducial method \eqref{beta_Press_Schechter} (solid blue line) with $\delta_2 = -3.05$ and $f_{\rm PBH} = 1$. For a comparison, we also provide extended mass functions based on several statistical uncertainties discussed in Appendix~\ref{Appd:PBH_statistics}. In all cases, one can see that $f(M_{\rm PBH}) < 10^{-6}$ for $M_{\rm PBH}/M_\odot \leq 10^{-18}$ so that the mass functions given in Figure~\ref{fig:f_k12} are compatible with current constraints\footnote{The plot of the observational constraints was generated using the publicly available code \href{https://github.com/bradkav/PBHbounds.}{\underline{PlotBounds}}.}. Notably, the green lines denote the Galactic bounds from: the {\sc Voyager1} data \cite{Boudaud:2018hqb}, the MeV diffuse flux observed by the {\sc Integral/Spi} detector \cite{Laha:2020ivk}, the 511 keV line observed in our Galaxy \cite{Laha:2019ssq, DeRocco:2019fjq} and the {\sc Comptel} measurements \cite{Coogan:2020tuf}. The red lines indicate the extragalactic bounds from {\sc Super-Kamiokande} \cite{Dasgupta:2019cae} and from the extragalactic background radiation \cite{Carr:2009jm}. Finally, the CMB and {\sc Edges} constraints are displayed in black.

There is a lower bound of $\delta_2$ that comes from the constant-mass approximation \eqref{appro_constant_mass} for the AD field, given that time-dependence appears in the Phase 3 mass eigenstates $m_{3\pm}$ when $\delta_2$ is away from the USR limit at $-3$. For example, with $m_\sigma/H_\ast = 1/2$, $N_\ast = 2$ and $N_{\rm end} = 20$, the constant-mass approximation $m_{3\pm} \approx m_\sigma$ holds for $\Lambda > 0.2 M_P$ at which the lower bound reads $\delta_2 > -3.12$. For $\Lambda = M_P$, the constant-mass holds for a larger parameter space with $\delta_2 > -3.3$. We remark that, however, the violation of constant-mass approximation does not imply a failure of the baryogenesis but just invokes an extension of the formalism in Section~\ref{Sec. massive_scalar_USR}.  

\subsubsection{Uncertainties}\label{Sec:uncertainty}
The PBH abundance from inflationary models of the USR class involves several uncertainties in the statistical approaches. The purpose of this section is to clarify the possible deviations from the fiducial parameter space led by various uncertain effects for the PBH formation. Here we summarize these uncertainties in terms of the difference in $N_\ast$ with respect to the fiducial value reported by $\beta_{\rm PS}$ defined in \eqref{beta_Press_Schechter}, where $N_\ast$ is obtained by $f_{\rm PBH}(\delta_2,N_\ast,N_{\rm end}) =1$ with fixed $\delta_2$ and $N_{\rm end}$.

\begin{itemize}
	\item \textit{Quantum diffusion.} The leading uncertainties for the PBH abundance in this scenario. In the exact USR limit ($\delta_2 = -3$), investigations in \cite{Biagetti:2021eep,Figueroa:2020jkf} suggest that the real abundance enhanced by quantum diffusion might lead to a deviation $\beta/\beta_{\rm PS} \sim 10^5 -10^{12}$, which can be translated into an $\mathcal{O}(10^{-2} - 10^{-1})$ deviation in $N_\ast$.  However, generalization of the method to the quasi-USR case with $\delta_2 < -3$ is required for the current scenario, where we expect a significant suppression of the effect of quantum diffusion due to the large effective inflaton mass of order $H_\ast$ (see Appendix~\ref{Appd_inflaton_mass}). As discussed in Section~\ref{Sec.CCP}, the viable parameter space for PBH as important dark matter with large enough baryon asymmetry allows a deviation $\beta/\beta_{\rm PS} \gg 10^{100}$ or namely an $\mathcal{O}(1)$ shift in $N_\ast$.
	
	\item \textit{Non-linear density.} The leading uncertainty for PBH formation arises from models of inflation. The deviation could be up to $\mathcal{O}(10^{-1})$ difference in $N_\ast$ (Appendix~\ref{Appd:NL_density}). The non-linear effect moves $N_\ast$ towards a larger value since the required spectral amplitude $A_{\rm PBH}$ becomes larger \cite{Young:2019yug}. The shape of the resulting mass function is similar to that of the critical collapse, see Figure~\ref{fig:f_k12}.
	
	\item \textit{Finite-size effect.} PBHs are not zero-size objects in real space and they only form at the local maximum (peaks) of the density contrast \cite{Bardeen:1985tr}. These spatial constraints introduce a $\nu_c^3$ factor in addition to the Press-Schechter method \cite{Wu:2020ilx,Young:2014ana}. As shown in Appendix~\ref{Appd:finite_size}, the effect gives a $\mathcal{O}(10^{-2})$ correction towards a smaller $N_\ast$ for a fixed PBH abundance. 
	
	\item \textit{Critical collapse.} The effect of critical collapse \cite{Niemeyer:1997mt,Yokoyama:1998xd,Musco:2012au,Musco:2008hv} is an intrinsic uncertainty for PBH formation from the reenter of large density perturbation into horizon. The extended mass function under critical collapse (Appendix~\ref{Appd:critical_collapse}) shows an evidently increased low-mass tail and a lowered peak value, yet it has a negligible correction of $\mathcal{O}(10^{-3})$ to the parameter $N_\ast$ for a fixed PBH abundance.   
\end{itemize}

Note that corrections from the transfer function that describes the time evolution of the density contrast after horizon reentry have been neglected in our results. There may be more uncertainties that arise from the choices of window functions, the profile dependence of the curvature perturbation, and the exact threshold value of gravitational collapse. We expect the combination of all uncertainties can only lead to a difference $\beta/\beta_{\rm PS} \ll 10^{100}$, which means at most an $\mathcal{O}(10^{-1})$ uncertainty for $N_\ast$.   

\subsection{The cosmic coincidence}\label{Sec.CCP}
The cold dark matter density today $\Omega_{\rm CDM0} =0.265$ in the standard $\Lambda$CDM Universe \cite{Aghanim:2018eyx} indicates that at matter-radiation equality $\Omega_{\rm CDMeq} = 0.42$ and the baryon density $\Omega_{\rm Beq} = m_B n_{\rm Beq} = 0.08$, where $m_B = 0.938$ GeV is the averaged nucleon mass and $n_{\rm Beq} = \vert Y_B\vert s(t_{\rm eq})$ is the baryon number density. The specific ratio between the two density parameters:
\begin{align}
 \Omega_{\rm CDMeq}/ \Omega_{\rm Beq} \approx 5,
\end{align}
is referred to as the ``cosmic coincidence problem.'' This cosmic coincidence also implies that $ \Omega_{\rm PBHeq}/ \Omega_{\rm Beq} \sim \mathcal{O}(1)$ if PBHs occupy $10$ $-$ $100$ percent of the dark matter density.

 Taking $H_0 = 67.36$ km s$^{-1}$ Mpc$^{-1}$, $\Omega_{\Lambda 0} = 1 -\Omega_{m0} = 0.6847$, and the redshift $z_{\rm eq} = 3402$ from \cite{Aghanim:2018eyx}, we find that $H_{\rm eq} = H_0 \sqrt{\Omega_{\Lambda0} + 2 \Omega_{m0}(a_0/a_{\rm eq})^3} \approx 2.264 \times 10^{-37}$ GeV and thus the baryon asymmetry at equality is $\vert Y_B\vert = 6.25 \times 10^{-11}$. Based on this estimate, we seek a viable parameter space for baryogenesis to satisfy $\vert Y_B\vert \gtrsim 10^{-10}$.

\begin{figure}
	\begin{center}
		\includegraphics[width=15cm]{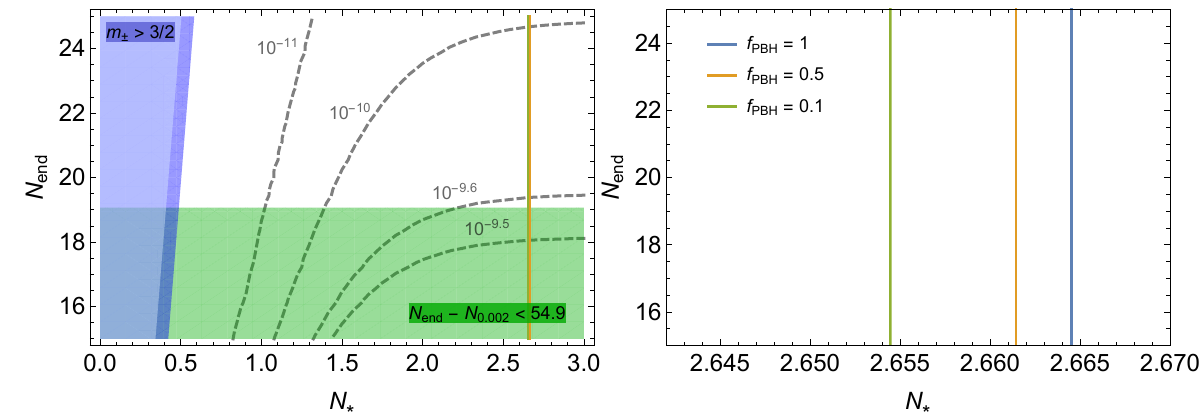} 
		%\hfill
		%\includegraphics[width=7cm]{CCP_zoom.pdf} 
		%\par
	\end{center}
	\caption{
		The parameter scan of PBH-baryon production from USR inflation with $\delta_2 = -3.05$ in the $\{N_\ast, N_{\rm end}\}$ plane, where $N = 0$ is chosen at the onset of USR transition in Figure~\ref{fig:USR_spectrum}. The PBH-to-dark matter ratio $f_{\rm PBH}(N_\ast, N_{\rm end}) = 0.1$, $0.5$, $1$ is estimated by the monochromatic relation $M_H = M_{\rm PBH}$. Initial conditions for $\sigma_{\pm}(N_\ast, N_{\rm end})$ are computed by \eqref{late_time_Phase3} based on the constant-mass approximation with $\{c_1, c_2, c_3 \} = \{-2, 1, 1 \}$, $m_\sigma = H_\ast/2 $, and $\Lambda = 0.3 M_P$. The contours are the final baryon asymmetry $\vert Y_B\vert$ resulting from the given initial conditions $\{N_\ast, N_{\rm end}\}$, where $m_I = 20H_\ast$ and $\Gamma_I = 10^{13}$ GeV are taken in the reheating scenario. Colored regions in the left panel are excluded by the minimal $e$-fold number of single-field inflation and the condition $m_\pm/H_\ast < 3/2$ for the late-time approximation of \eqref{late_time_Phase3}. Right Panel: A zoom-in for the ratio $f_{\rm PBH}(N_\ast, N_{\rm end})$ on the $N_\ast$ axis.  
		\label{fig:CCP}}
\end{figure}

In Figure~\ref{fig:CCP}, we scan the parameter space $\{N_\ast, N_{\rm end}\}$ for the USR inflation with a slightly tilted spectrum $\delta_2 = -3.05$ that can be realized $f_{\rm PBH} > 0.1$ and keep the constant-mass formalism \eqref{appro_constant_mass} as a good approximation. The results in Figure~\ref{fig:CCP} show that baryogenesis triggered by the USR inflation with $\vert Y_B\vert \gtrsim 10^{-10}$ is compatible with PBH dark matter in the range of $19 < N_{\rm end} < 24$ and this viable parameter space is not sensitive to a small variation in $N_\ast$. On the other hand, the PBH abundance is very sensitive to a perturbation in $N_\ast$ so that a specific value in the range of $0.1 \leq f_{\rm PBH} \leq 1$ will fix a precise value of $N_\ast$ in the region of $N_\ast > 2$. In other words, if PBHs compose an important dark matter fraction with $0.1 \leq f_{\rm PBH} \leq 1$, the inflationary parameter $N_\ast$ is fixed to the space that gives $\vert Y_B\vert \gtrsim 10^{-10}$ so that $\Omega_{\rm PBHeq}/ \Omega_{\rm Beq} \sim \mathcal{O}(1)$ becomes a natural outcome in the scenario. This provides a potential explanation to the cosmic coincidence problem for  PBH dark matter. 

The above conclusion is unchanged even if one takes into account the effects of critical collapse (\ref{Appd:critical_collapse}) or the non-linear density corrections (\ref{Appd:NL_density}) in the PBH mass function. As one can see from Figure~\ref{fig:f_k12}, the effect of critical collapse slightly decreases the peak amplitude of $f(M_{\rm PBH})$ and lifts the population of PBHs towards the small mass limit. As a result, the values of $N_\ast$ required by $0.1 \leq f_{\rm PBH} \leq 1$ are slightly larger than those without critical collapse. Similarly, the non-linearity of the density contrast asks a larger $N_\ast$ by $\mathcal{O}(0.1)$, yet a constant shift of $N_\ast$ towards the region of $N_\ast > 2$ does not affect the resulting $\vert Y_B\vert $ in this scenario.

Note that in the paradigm of slow-roll single-field inflation, the exclusion of a perfect scale-invariant $P_\zeta$ in general requires a finite $e$-fold number $50 < N_{\rm end} - N_{0.002} < 60$, where $N_{0.002}$ is the number at the pivot scale of CMB measurements $k_{\rm CMB} = 0.002$ Mpc$^{-1}$ \cite{Akrami:2018odb}. The non-observation of spatial curvature requires an additional $4.9$ $e$-folds to produce the pivot scale \cite{Akrami:2018odb}. These two constraints imply a minimal $e$-fold number from CMB scales to the smallest scales at the end of inflation: $\ln (k_{\rm end}/k_{\rm CMB}) \approx \ln(a_{\rm end}/a_{\rm CMB})> 54.9$,
which translates into a constraint $N_{\rm end} = \ln(k_{\rm end}/k_0) > 54.9 - \ln (k_{0}/k_{\rm CMB})$. 

Before closing this section, we remark that the final baryon asymmetry triggered by the USR transition during inflation always approaches  a constant value in the regime of $N_\ast > 2$ for a heavy AD field with $m_\sigma \sim \mathcal{O}(H_\ast)$, which we have numerically confirmed with various choices of the $\mathcal{O}(1)$ parameters, $c_1$, $c_2$, $c_3$ (but avoided the fine cancellation between $c_1$ and $c_2$), and $m_I > 10 H_\ast$. This conclusion relies on the constant-mass approximation \eqref{appro_constant_mass} which sets a lower bound to the cutoff as $\Lambda > \sqrt{\epsilon_{\rm CMB}}M_P H_\ast/m_\sigma$. However, a $\Lambda$ below this bound does not necessarily mean a failure of the baryogenesis, but only requires a modification of the solutions in Section~\ref{Sec. massive_scalar_USR} to that of a time-dependent mass term. In general, a larger $Y_B$ is obtained with a lower $\Lambda$ since it gives a larger source of the baryon number violation in the equation of motion (namely the $c_2$ term).

The ``asymptotic constant'' behavior of $\vert Y_B \vert$ implies that the viable parameter space for explaining the cosmic coincidence problem holds as long as $A_{\rm PBH}/A_{\rm CMB} > 5\times 10^2$ for $\delta_2 \lesssim -3$, which can be translated into the viable range $10^{-6} < A_{\rm PBH} < 10^{-1}$ with the upper bound set by the perturbativity of the inflation model.\footnote{For a reference in terms of $f_{\rm PBH}$ based on the fiducial method, the value $N_\ast = 2.3$ with $\delta_2 = -3.05$ gives $f_{\rm PBH} \sim 10^{-130}$ and $A_{\rm PBH} \sim 3\times 10^{-3}$, and $N_\ast = 2.3$ with $\delta_2 = -3.1$ gives $f_{\rm PBH} \sim 10^{-83}$ and $A_{\rm PBH} \sim 5\times 10^{-3}$. }
As a result, even if there could be some $\mathcal{O}(1 -10)$ uncertainty for the necessary amplitude $A_{\rm PBH}$ to realize $0.1 \leq f_{\rm PBH} \leq 1$ due to any uncovered effect involved in the PBH formation from USR inflation,
% \cite{DeLuca:2019qsy,Atal:2018neu,Taoso:2021uvl,Germani:2019zez} 
the prediction of the ratio $\Omega_{\rm PBHeq}/ \Omega_{\rm Beq}$ remains unaffected.

\section{Conclusion}\label{Sec:conclusion}
The sharp deceleration of the slow-roll dynamics on small scales into an ``ultra-slow-roll (USR)'' phase is a generic mechanism to enhance the primordial power spectrum for PBH formation in single-field inflation. If PBHs indeed play an important role in dark matter, the cosmic coincidence problem among the energy densities of the Universe today might hint a correlated origin for baryon and dark matter. In this work, we have explored and confirmed the viability of baryogenesis, based on the AD mechanism, with modified initial conditions and baryon-number-violating operators triggered by the generic USR transition for PBH formation. 

Baryogenesis driven by the USR transition during inflation significantly extends the viable parameter space for the initial conditions from preceding conclusions. It does not rely on the presence of ``flat directions'' nor a tachyonic soft mass term induced by non-minimal Kahler couplings. Scalar fields with initial VEVs much smaller than the Hubble scale of inflation can create sufficient baryon asymmetry as long as they were dynamically excited during inflation. In other words, the AD mechanism of baryogenesis can be realized with a minimal Kahlar potential or without the assumption of supersymmetry. Albeit the formalism presented in this work has been focused on the constant-mass assumption for the AD field near the USR limit ($\delta_2 \simeq -3$) with a bound $0 < m_\pm/H_\ast < 3/2$ on the effective masses, we expect successful baryogenesis beyond those constraints. The coherent production of scalar motion away from the USR limit and the modified late-time behavior for $m_\pm/H_\ast > 3/2$ are interesting topics for further investigations.

The asymptotic constant behavior of the final baryon asymmetry towards the large USR duration limit ($N_\ast \gg 1$) is a generic feature of the presented scenario and this fact has important implications to the cosmic coincidence problem. For PBHs to occupy a significant fraction of the dark matter density $\Omega_{\rm PBH}$, the value of $N_\ast$ must be precisely fixed, whereas the exact value depends on various uncertainties involved in the statistics of the PBH abundance. However, all statistical methods tested in this work suggest that the allowed space of $N_\ast$ for PBH dark matter lies deep inside the constant plateau of the correct baryon asymmetry, where the coincidence between $\Omega_{\rm PBHeq}$ and $\Omega_{\rm Beq}$ can incorporate at least $10^{100}$ orders of uncertainties in the ratio of $f_{\rm PBH} \equiv \Omega_{\rm PBH}/\Omega_{\rm DM}$. Nevertheless, despite the fact that the USR transition provides a quantitative connection between baryon and PBH densities with a viable answer to the cosmic coincidence problem independent of the statistical uncertainties, the existing fine-tuning problem for the inflaton potential to realize PBH dark matter inevitably enters the story of cosmology. On the other hand, the asteroid-mass window for PBH dark matter, one of the fundamental assumption in this work, could be tested by near future astrophysical or gravitational-wave experiments \cite{Nemiroff:1995ak, Katz:2018zrn, Jung:2019fcs,Ballesteros:2019exr, Coogan:2020tuf,Capela:2013yf,Graham:2015apa,Ray:2021mxu,Dutta:2020lqc,Carr:2020gox,Green:2020jor,Ali-Haimoud:2019khd,Bai:2018bej}.    

\acknowledgments
%This is the most common positions for acknowledgments. A macro is available to maintain the same layout and spelling of the heading.
We thank Julien Lesgourgues and Vincent Vennin for helpful discussions. 
Y.-P. Wu and K.~Petraki were supported by the Agence Nationale de la Recherche (ANR) Accueil de Chercheurs de Haut Niveau (ACHN) 2015 grant (“TheIntricateDark” project). 
K.~Petraki was also supported by the NWO Vidi grant ``Self-interacting asymmetric dark matter.''
%E. Pinetti acknowledges a grant from the Université Franco-Italienne under Bando Vinci 2020.
E. Pinetti is supported by: the Fermi Research Alliance, LLC under Contract No. DE-AC02-07CH11359 with the U.S. Department of Energy, Office of High Energy Physics; Department of Excellent grant 2018-2022, awarded by the Italian Ministry of Education, University and Research (MIUR); Research grant of the Italo-French University, under Bando Vinci 2020.
The project has received funding from the European Union’s Horizon 2020 research and innovation programme under grant agreement No 101002846 (ERC CoG ``CosmoChart'').
%\paragraph{Note added.} This is also a good position for notes added after the paper has been written.

\appendix
%\section{Some title}
%Please always give a title also for appendices.
\section{The inflaton mass in quasi-USR inflation}\label{Appd_inflaton_mass}
The exact USR limit with $\delta_2 = -3$ corresponds to a region of completely flat potential in which the inflaton field $\phi$ is effectively massless. In this region stochastic effect driven by modes well inside the horizon can play an important role in the VEV of $\phi$, leading to non-linear translation of the curvature perturbation $\zeta$ from $\phi$ beyond the standard Gaussian approximation \cite{Biagetti:2018pjj,Ezquiaga:2019ftu,Pattison:2021oen,Figueroa:2020jkf,Biagetti:2021eep}. In this section, we drive an upper bound for $\delta_2$ to prevent the breakdown of the Gaussian assumption for $\zeta$.

The constant-rate condition $\dot{\delta}_2 = 0$ used for the USR template in Section~\ref{Sec. USR_template} is in fact a constraint on the inflaton mass. To see this, one uses $\delta = \ddot{\phi}/(H\dot{\phi})$ to replace the classical motion $\ddot{\phi} + 3H\dot{\phi} + V_{\phi} = 0$ to obtain $\delta+ 3 = -V_\phi/(H\dot{\phi})$. Taking the time derivative of $\delta$ one obtains the relation with the inflaton mass $V_{\phi\phi}$ \cite{Ng:2021hll}, where the constant-rate condition $\dot{\delta} = 0$ gives 
\begin{align}\label{inflaton_mass}
\frac{V_{\phi\phi}}{H^2} = \left(\epsilon_H - \delta\right)\left(\delta + 3\right),
\end{align}
where $\epsilon_H \rightarrow 0$ manifests a de Sitter background.
One can see that $\delta = 0$ (slow-roll) and $\delta = -3$ (USR) are two special cases for the massless inflaton. The condition $\delta_3 = -\delta_2 -3$ with $\delta_2 \leq -3$ implies a continuous inflaton mass from Phase 2 to 3 so that the scaling power of $P_\zeta$ is also continuous \cite{Ng:2021hll}.

To avoid a large quantum diffusion led by the stochastic fluctuation of $\phi$, we ask the effective mass $m_\phi^2 \equiv \vert V_{\phi\phi} \vert$ to satisfy the condition, $m_\phi^2/H > H/(2\pi)$, so that the classical deviation is larger than the size of the quantum fluctuation. With $V_{\phi\phi} = -\delta(\delta + 3) H^2$, one finds that $m_\phi/H > 1/\sqrt{2\pi} $ gives the upper bound $\delta_2 < -3.05$.  

The statistics of PBH abundance in the exact USR inflation with $\delta_2 = -3$ is very sensitive to the effect of quantum diffusion \cite{Ezquiaga:2019ftu,Pattison:2021oen}. For more physical cases with $\delta_2 < -3$, the inflaton mass \eqref{inflaton_mass} led by the constant-rate condition $\dot{\delta}_2 = 0$ in fact has fixed solutions of the inflaton mode functions (with the de Sitter condition $\epsilon_H \rightarrow 0$), see \cite{Ng:2021hll}. In other words, power spectrum of the Gaussian inflaton perturbation $\delta\phi$ has been uniquely determined by the background evolution of $\phi$, where in the USR limit with $\delta_2 \rightarrow -3$ the spectrum $P_{\delta\phi}$ is exactly scale invariant. There is no degree of freedom to impose an additional delta-like spectrum for $\delta\phi $ as in \cite{Biagetti:2021eep} and the effect of quantum diffusion in the probability distribution function of the density contrast away from the USR limit is left for future effort. 

\section{Baryogenesis from massless AD field}\label{Appd_massless_AD}
Let us explore in this section the baryogenesis during inflation triggered by a charged scalar with a negligible bare mass. 
In the limit of $m_\sigma \rightarrow 0$, $m_{1\pm}$ and $m_{3\pm}$ given in \eqref{appro_constant_mass} are still constants yet the scalar mass in Phase 2 only given by the $c_3$ term becomes decaying with time, where
\begin{align}\label{appro_massless}
\nonumber
m_{1\pm}^2 =& -3 \frac{c_1 \pm c_2}{\Lambda} \sqrt{2\epsilon_{\rm CMB}} M_PH_\ast^2 \\
m_{2\pm}^2 =& \frac{c_3}{\Lambda^2} 2\epsilon_{\rm CMB}M_P^2 H_\ast^2 e^{2\delta_2N} = \bar{m}^2 e^{-6 N}, \\
m_{3\pm}^2 =& -3 \frac{c_1 \pm c_2}{\Lambda} \sqrt{2\epsilon_{\ast}} M_PH_\ast^2
+ \frac{c_3}{\Lambda^2} 2\epsilon_{\ast} M_P^2H_\ast^2,
\end{align}
and we denote $\bar{m}^2 = 2 c_3 \epsilon_{\rm CMB} M_P^2 H_\ast^2/\Lambda^2$ for convenience. To study the baryogenesis in this limit, the coefficients $c_i$ are chosen such that $m_{i\pm}$ in each phase is always positive defined. We ask $m_{1\pm} \sim \mathcal{O}(H_\ast)$ for the AD field to be compatible with the isocurvature constraint in Phase 1, which gives an upper bound to the cutoff as $\Lambda < 3\sqrt{2\epsilon_{\rm CMB}}M_P$.

For $m_\sigma \ll H_\ast$, we adopt the massless approximation \eqref{appro_massless} to the mass eigenstates, where $m_{2\pm}$ are degenerated but they have explicit time-dependence. The Phase 2 equation of motion for $\sigma_{\pm}$ becomes
\begin{align}
\ddot{\sigma}_{2\pm} + 3 H_\ast \dot{\sigma}_{2\pm} + \bar{m}^2e^{2\delta_2 H_\ast t} \,\sigma_{2\pm} = 0, \qquad t_0\leq t < t_\ast,
\end{align}
where $\bar{m}^2 = 2 c_3 \epsilon_{\rm CMB} M_P^2 H_\ast^2/\Lambda^2$, and in the USR limit $\delta_2 = -3$ the solution takes a simple form of
\begin{align}
\sigma_{2\pm}(N) = \sqrt{\frac{3}{8\pi^2}} \frac{H_\ast^2}{m_{1\pm}} 
\cos \left[\frac{\bar{m}}{3H_\ast} \left(1-e^{-3N}\right)\right].
\end{align}
Note that initial conditions given by \eqref{initial_condition_Phase1} at $N =0$ have been used.
The decaying-mass solutions in Phase 2 modify the coefficients of the constant-mass Phase 3 solutions $\sigma_{3\pm} = C_\pm  e^{-\Delta_{3\pm}^+ N} + D_\pm e^{-\Delta_{3\pm}^- N}$, by matching boundary conditions at $N = N_\ast$, as
\begin{align}
C_\pm = \frac{e^{\Delta_{3\pm}^+ N_\ast}}{\Delta_{3\pm}^- - \Delta_{3\pm}^+} \left(\sigma_{\ast\pm} \Delta_{3\pm}^- - A_\pm\right),\quad
D_\pm = \frac{e^{\Delta_{3\pm}^- N_\ast}}{\Delta_{3\pm}^- - \Delta_{3\pm}^+} \left(A_\pm -\sigma_{\ast\pm} \Delta_{3\pm}^+  \right),
\end{align}
where $\sigma_{\ast\pm} = \sigma_{2\pm}(N_\ast)$,
$A_\pm = \sigma_{0\pm} (\bar{m}/H_\ast) e^{-3N_\ast} \sin[\bar{m}(1-e^{-3N_\ast})/(3H_\ast)]$ and $\sigma_{0\pm} \equiv \sigma_\pm(t_0) = \sqrt{3/(8\pi^2)}  H_\ast^2/m_{1\pm}$.

\begin{figure}
	\begin{center}
		\includegraphics[width=7cm]{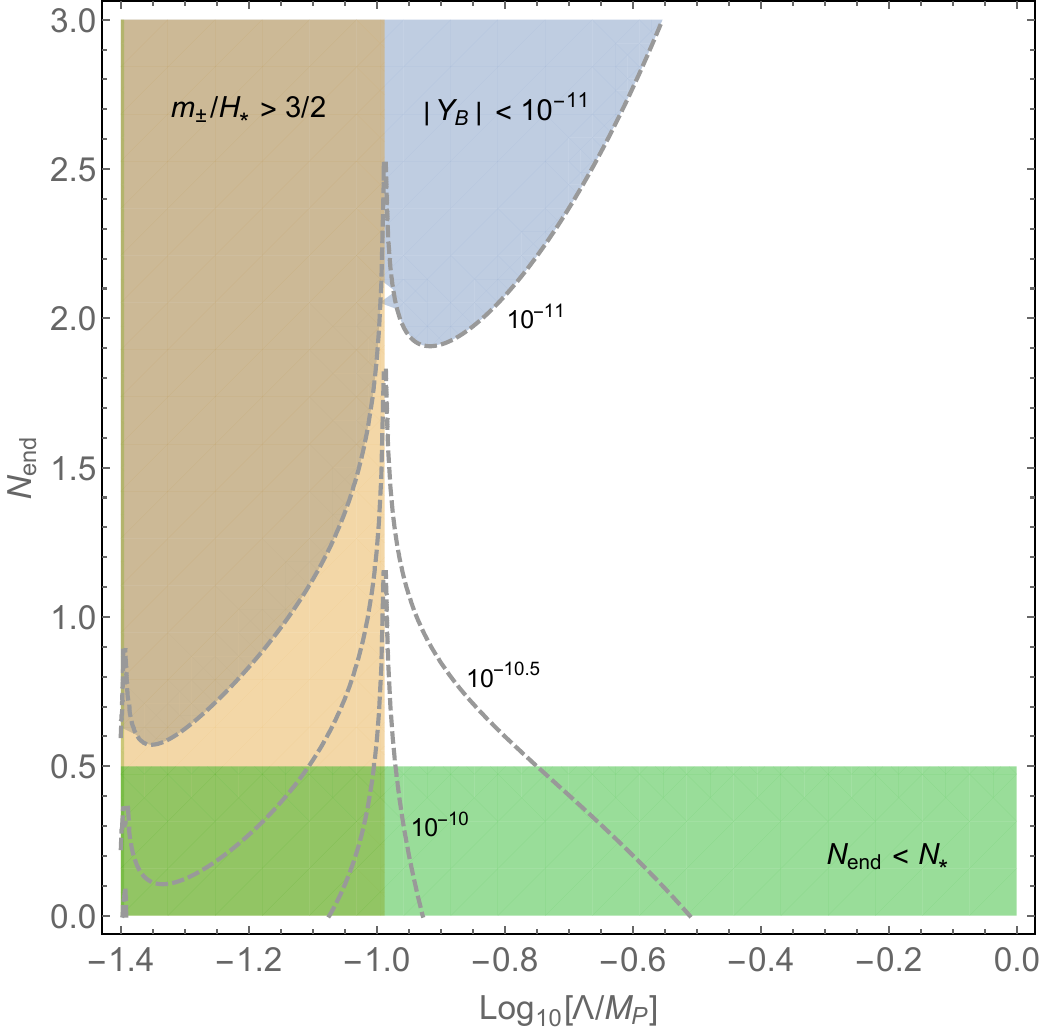} 
		%\par
	\end{center}
	\caption{The parameter scan for the baryon asymmetry $\vert Y_B \vert$ at the end of inflation from massless AD field ($m_\sigma = 0$) in the USR limit $\delta_2 = -3$ with $\{c_1, c_2, c_3 \} = \{-2, 1, 1 \}$, $N_\ast = 0.5$ and $H_\ast = 2.37 \times 10^{13}$ GeV. Dashed lines are contours of $\vert Y_B\vert$. Colored regions are excluded by validity of the late-time approximation for $m_{3\pm}/H_\ast < 3/2$ and the existence of the USR phase ($N_{\rm end} > N_\ast$). \label{fig:Y_B_massless AD}}
\end{figure}

The resulting baryon asymmetry estimated at the end of inflation from a massless AD field ($m_\sigma \rightarrow 0$) can marginally reach the observational value $\vert Y_B\vert \lesssim 10^{-10}$ in a very limited parameter space. An example using the late-time approximation \eqref{Y_B__analyitc_inflation_end} for $m_\pm/H_\ast < 3/2$ with $N_\ast = 0.5$ and $H_\ast = 2.37 \times 10^{13}$ GeV is given in Figure~\ref{fig:Y_B_massless AD}. By definition, the ending time of inflation $N_{\rm end} > N_\ast$ is necessary for the existence of the USR phase (Phase 2).

\section{The statistics of PBH abundance}\label{Appd:PBH_statistics}
In this section we provide a more detailed formulation of the uncertain effects in the statistics of PBH abundance considered in Section~\ref{Sec:uncertainty}.  In Table~\ref{Table_Ns}, a summary is given about the condition $f_{\rm PBH} = 1$ for PBH as all dark matter from different statistical approaches in terms of the duration of the USR phase $N_\ast$ and the maximal spectral amplitude $A_{\rm PBH} = A_{\rm CMB} e^{-N_\ast(6+4\delta_2)}$. The corresponding mass function in each approach given in Figure~\ref{fig:f_k12} is compatible with existing observational constraints from galactic \cite{Boudaud:2018hqb,DeRocco:2019fjq,Laha:2019ssq,Laha:2020ivk}, extra-galactic \cite{Carr:2009jm,Dasgupta:2019cae}, and CMB or 21cm experiments \cite{Poulin:2016anj,Clark:2016nst,Clark:2018ghm,Hektor:2018qqw,Mittal:2021egv}.

\begin{table}[t]
	\centering
	\begin{tabular}{| p{2.5cm} | p{2cm} | p{2cm}| p{2cm} | p{2cm} |}
		\hline
		& $\beta_{\rm PS}$ & $\beta_c$ & $\beta_{\rm 3d}$ & $\beta_{\rm NL}$ \\
		\hline
		$\delta_2 = -3.05$ & 2.6645 & 2.6687 & 2.6423 & 2.7473 \\
		\hline
		$A_{\rm PBH}$ & 0.0329 & 0.0286 & 0.0337& 0.0549 \\
		\hline
		$\delta_2 = -3.1$ & 2.5837 & 2.5877 & 2.5628 & 2.6638 \\
		\hline
		$A_{\rm PBH}$ & 0.0334 & 0.0292 & 0.0342 & 0.0558 \\
		\hline
	\end{tabular}
\caption{The duration of the USR phase, $N_\ast$, and the corresponding spectral amplitude, $A_{\rm PBH}$, for realizing PBH as all dark matter $f_{\rm PBH} = 1$ from various statistical approaches with $N_{\rm end} = 20$.\label{Table_Ns}}
\end{table}

\subsection{The fiducial approach}\label{Appd:PS_method}
In this work, we forecast the PBH abundance by assuming the monochromatic relation $M_{\rm PBH} = M_H$ (no critical collapse)\footnote{Due to the effect of critical collapse \cite{Niemeyer:1997mt,Yokoyama:1998xd,Musco:2012au,Musco:2008hv}, $M_{\rm PBH}$ in general can be much smaller than the horizon mass $M_H$ at the epoch of formation.}
 at each comoving epoch so that the ratio $M_{\rm PBH}/M_H = 1$ in \eqref{PBH_density_general}. Assuming that $F$ is the only variable relevant to the PBH formation without further constraints, we have $f_c(y_i) = 1$ and $P(F,y_i,\sigma_{Fi}) = e^{F^2/(2\sigma_{F0}^2)}/\sqrt{2\pi}$ so that \eqref{PBH_density_general} simply gives the Press-Schechter formula \eqref{beta_Press_Schechter} (up to a factor of 2). 

Conventionally, the Gaussian field $F$ is identified as the density perturbation of radiation converted from the decay of inflaton during reheating.
%\footnote{The curvature perturbation $\zeta$ is another natural option for the fundamental random field in the statistics of PBH abundance from models of inflation \cite{Yoo:2018esr,DeLuca:2019qsy,Kalaja:2019uju}.}
At linear order, the power spectrum of $F$ is related to $P_\zeta$ according to
\begin{align}\label{P_F}
P_F =  \frac{4(1+w)^2}{(5+3w)^2} \left(\frac{k}{a H}\right)^4 P_\zeta,
\end{align}
where $w = 1/3$ is the equation of state of the Universe in radiation domination.
The linear simplification \eqref{P_F} of the density spectrum inevitably loses non-linear and non-Gaussian contributions from $P_\zeta$ \cite{DeLuca:2019qsy,Kalaja:2019uju,Kawasaki:2019mbl,Young:2019yug,Saito:2008em,Byrnes:2012yx,Tada:2015noa,Franciolini:2018vbk,Atal:2018neu,Taoso:2021uvl,Germani:2019zez,Young:2015cyn}. However, given that the PBH abundance is exponentially sensitive to the peak value $\nu$, the statistical results based on non-linear or non-Gaussian approaches can usually be mimicked by the linear Gaussian statistics with a mild change in the input parameters (or, to be more specifically, the spectral amplitude $A_{\rm PBH}$), see also \cite{DeLuca:2019qsy,Taoso:2021uvl}.  

The gain from using the linear relation \eqref{P_F} is that one can obtain explicit analytic expressions for the spectral moments \eqref{spectral_moment} given by the USR template \eqref{CR_template}. Importing the Gaussian window function $W(kR) = \exp[-k^2R^2/2]$, one can decompose the template \eqref{CR_template} into one blue-tilted trapezoidal spectrum (from $k_{\rm min}$ to $k_0$) plus one step/trapezoidal spectrum (from $k_0$ to $k_{\rm end}$) to obtain the analytic result of $\sigma_{Fi}$ based on the method in \cite{Wu:2020ilx}, and it reads
\begin{align}
\sigma_{Fi}^2 (R_H) &= \frac{4(1+w)^2}{(5+3w)^2} \frac{A_{\rm PBH}}{2 R_H^{2i}} \\\nonumber
\times&
\left[ \frac{\Gamma(5+i +\delta_2, r_0^2)}{r_0^{6+2\delta}} -  \frac{\Gamma(5+i +\delta_2, r_{\rm end}^2)}{r_0^{6+2\delta}} + \frac{\Gamma(4+i, r_{\rm min}^2)}{r_0^4} - \frac{\Gamma(4+i, r_0^2)}{r_0^4}\right],
\end{align}
where $r_0 = k_0 R_H$, $r_{\rm min} = k_{\rm min} R_H$ and $r_{\rm end} = k_{\rm end} R_H$. Note that $A_{\rm PBH} = A_{\rm CMB} e^{-N_\ast(6+4\delta_2)}$, $k_{\rm min} = (A_{\rm CMB}/A_{\rm PBH})^{1/4} k_0$, and $k_{\rm end} = k_0e^{N_{\rm end}}$ are all expressed by free parameters $\{\delta_2, N_\ast, N_{\rm end}\}$ of the USR inflation. $k_0$ is fixed by the targeting mass range of PBHs.

%\footnote{The Press-Schechter approach given by \eqref{beta_Press_Schechter} is based on an one-variable Gaussian statistics with respect to the peak value $\nu$. To take into account the spatial constraints for PBHs being 3-dimensional objects in real space, one shall construct multi-variable statistics with respect to $\nu$ and its spatial derivatives \cite{Bardeen:1985tr,Green:2004wb,Young:2014ana,Wu:2020ilx,Suyama:2019npc,Young:2020xmk}. However, due to the fact that $P_\zeta$ considered in this work is subject to board power spectrum spanning a wide mass range, the difference with or without spatial constraints in the final PBH abundance can be negligible when comparing with other uncertainties, see \cite{Young:2014ana,Wu:2020ilx}. }

\subsection{The effect of critical collapse}\label{Appd:critical_collapse}
In this section, we derive the PBH mass function with extended $M_{\rm PBH}$ to $M_H$ relation led by the effect of critical collapse \cite{Niemeyer:1997mt,Yokoyama:1998xd,Musco:2012au,Musco:2008hv}. This effect is conventionally described by the scaling relation of the form:
\begin{align}\label{mPBH_critical_collapse}
M_{\rm PBH} = K M_H (F - F_c)^{\gamma_c},
\end{align}
where $K = 3.3$ and $\gamma_c = 0.35$ are two numerical constants. $F_c$ is the threshold density perturbation for gravitational collapse. The scaling formula generalizes the peak value $\nu = F/\sigma_{F0}$ to be a function of both $M_{\rm PBH}$ and $M_H$ so that \cite{Wu:2020ilx,Byrnes:2018clq}:
\begin{align}
\beta_c\left(M_{\rm PBH}, M_H\right) = \frac{K}{\sqrt{2\pi} \gamma_c \sigma_{F0}(M_H)} \left(\frac{M_{\rm PBH}}{K M_H}\right)^{1+1/\gamma_c} 
\exp\left[-\frac{F^2}{2\sigma_{F0}^2(M_H)}\right],
\end{align}
where $F = [ M_{\rm PBH}/(K M_H) ]^{1/\gamma_c} + F_c$.

The PBH mass function based on the extended mass relation at matter-radiation equality is therefore a sum of the contribution at each epoch of the horizon mass $M_H$, where
\begin{align}\label{mass_function_critical_collapse}
f_c(M_{\rm PBH}) = \frac{1}{\Omega_{\rm DM,eq}} \int_{\ln M_{\rm min}}^{\ln M_{\rm eq}} \left(\frac{M_{\rm eq}}{M_H}\right)^{1/2} 
\beta_c\left(M_{\rm PBH}, M_H\right) d \ln M_H.
\end{align}
The lower limit $M_{\rm min}$ comes from the upper bound for $F$ as over-large density perturbation would have already collapsed to form PBHs in earlier epoch. We adopt a conservative upper bound $F_{\rm max} \leq 2 F_c$ so that one can derive the lower limit $M_{\rm min} \geq M_{\rm PBH}/(KF_c^{\gamma_c})$ from the scaling relation \eqref{mPBH_critical_collapse}.

Finally, the PBH ratio in dark matter is the sum of all masses at equality: $f_{\rm PBH} \equiv \int f_c(M_{\rm PBH}) d\ln M_{\rm PBH}$. As a comparison, for $f_{\rm PBH} = 1$ with $k_0 = 7.3\times 10^{12}$ Mpc$^{-1}$, we find that $N_\ast = 2.6687$ ($2.6645$) with (or without) the effect of critical collapse.

\subsection{The finite-size effect}\label{Appd:finite_size}
% Fixing $f_{\rm PBH} = 1$, a narrow mass distribution with $\delta = -3.2$ shows $N_\ast = 2.449$, and with $\delta = -3.1$ we find $N_\ast = 2.584$. For a comparison, we also adopt Gaussian peak statistics with 3-dimensional spatial constraints defined in \cite{Bardeen:1985tr}. 
%In the high peak limit $\nu_c = \Delta_c/\sigma_0 \gg 1$, the peak statistics \cite{Bardeen:1985tr} gives $\beta_{\rm pk}(M_H) \simeq \frac{1}{2}Q^{3/2}\nu_c^3\beta_{\rm PS}$ \cite{Wu:2020ilx,Young:2014ana}, where $Q\equiv R^2\sigma_1^2/(3\sigma_0^2)$ is $\mathcal{O}(1)$ near $k_0$ and $\nu_c^3$ is the finite-size effect for PBHs in 3 dimensional real space. For $f_{\rm PBH} = 1$ with fixed $\delta$, the differences of the resulting $N_\ast$ between $\beta_{\rm PS}$ and $\beta_{\rm pk}$ is $\mathcal{O}(10^{-2})$. 

The fact that PBHs are not point particles in real space and they only form at the local maximum of the density contrast introduces spatial constraints of the peak value $\nu = F/\sigma_{F0}$ with respect to its first and second spatial derivatives. These constraints in 3-dimensional space lead to non-trivial functions $f_c(y_i)$ among the spectral moments in the PBH density \eqref{PBH_density_general} and invoke a system of 10-variable Gaussian statistics \cite{Bardeen:1985tr}. In the high-peak limit $\nu_c = F_c/\sigma_{F0} \gg 1$, the mass fraction with 3-dimensional spatial constraints is found as \cite{Wu:2020ilx,Young:2014ana}:
\begin{align}
\beta_{\rm 3d}(M_H) \simeq \frac{1}{2}Q^{3/2}\nu_c^3\beta_{\rm PS}(M_H),
\end{align}
where $Q\equiv R^2\sigma_{F1}^2/(3\sigma_{F0}^2)$ is an $\mathcal{O}(1)$ factor near the pivot scale $k_0$ of the USR transition, and $\nu_c^3$ is nothing but the finite-size effect for PBHs in 3-dimensional real space. The detailed derivation of $\beta_{\rm 3d}$ and the extended version with the effective of critical collapse can be found in \cite{Wu:2020ilx}.

The finite-size effect usually becomes negligible if $P_F$ is a broad spectrum spanning a wide mass range.
By fixing $f_{\rm PBH}(\delta_2,N_\ast,N_{\rm end}) = 1$ at $N_{\rm end} = 20$ for a broad spectrum with $\delta_2 = -3.05$, we find that the result from $\beta_{\rm 3d}$ ($\beta_{\rm PS}$) gives $N_\ast = 2.6423$ ($2.6645$). For a comparison, a narrower spectrum with $\delta_2 = -3.1$ at $N_{\rm end} = 20$, we find that $N_\ast = 2.563$ from $\beta_{\rm 3d}$ and $N_\ast = 2.584$ from $\beta_{\rm PS}$. 

\subsection{Non-linear density contrast}\label{Appd:NL_density}
The general solution of the Einstein equation for the relation between the curvature perturbation and the density contrast of radiation is non-linear. Recent studies \cite{Musco:2018rwt,Young:2019yug,Harada:2015yda} suggested that the density perturbation of radiation, $\delta\rho_r/\rho_r$, in terms of the compaction function provides a well-defined criterion for PBH formation, which is related to the profile of the curvature perturbation, $\zeta(r)$, through the non-linear equation as
\begin{align}
\frac{\delta\rho_r}{\rho_r} \equiv F_{NL} = F - \frac{3}{8}F^2,
\end{align}
where $F = -4r_m \partial_r \zeta(r_m)$ is the linear density contrast imposed in the fiducial approach and $r_m$ is the local maximum of the compaction function. Such a non-linear relation promotes the extended mass relation of critical collapse \eqref{mPBH_critical_collapse} to be
\begin{align}\label{mPBH_non_linear}
M_{\rm PBH} = K M_H (F_{\rm NL} - F_c)^{\gamma_c} = K M_H \left(F-\frac{3}{8}F^2 - F_c \right)^{\gamma_c}.
\end{align}
Completing the square of the linear perturbation (with $2/3 > F > F_c$ \cite{Young:2019yug}), one can derive the modified relation
\begin{align}\label{dF_non_linear}
dF = \frac{4}{3\gamma_c} \left[\frac{16}{9}- \frac{8}{3}F_c -\frac{8}{3}\left(\frac{M_{\rm PBH}}{K M_H}\right)^{1/\gamma_c} \right]^{-1/2} 
\left(\frac{M_{\rm PBH}}{K M_H}\right)^{1/\gamma_c} d\ln M_{\rm PBH}.
\end{align}
Given that our goal is only to clarify the effect of non-linearity for the mass function based on the fiducial approach, we compute the PBH density by using the one-variable formula with Gaussian probability distribution function as
\begin{align}
\Omega_{\rm PBH}(M_H) = \frac{1}{\sqrt{2\pi} \sigma_{F0}} \int_{F_c} \frac{M_{\rm PBH}}{M_H} e^{-F^2/(2\sigma_{F0}^2)} dF.
\end{align}
Taking the modified relations from \eqref{mPBH_non_linear} and \eqref{dF_non_linear}, one can derive the mass fraction from non-linear density contrast of the form
\begin{align}
\beta_{\rm NL} (M_{\rm PBH}, M_H)= 
&\frac{4K}{3\sqrt{2\pi} \gamma_c \sigma_{F0}} \left(\frac{M_{\rm PBH}}{K M_H}\right)^{1+1/\gamma_c}\\\nonumber
&\times\left[\frac{16}{9}- \frac{8}{3}F_c -\frac{8}{3}\left(\frac{M_{\rm PBH}}{K M_H}\right)^{1/\gamma_c} \right]^{-1/2} 
\exp\left[-\frac{F^2}{2\sigma_{F0}^2}\right].
\end{align}
The unified result of non-linear density contrast with the finite-size effect of density peaks is provided in \cite{Young:2019yug}.

The extended mass function from non-linear density perturbation can be computed similarly as \eqref{mass_function_critical_collapse} by using $\beta_{\rm NL}$.
Comparing with the resulting $N_\ast$ for $f_{\rm PBH}(\delta_2,N_\ast,N_{\rm end}) = 1$ with fixed $\delta_2$ and $N_{\rm end}$ based on the fiducial method, the mass function from non-linear density contrast asks a larger $N_\ast$ by $\mathcal{O}(10^{-1})$. 
%For example, at $N_{\rm end} =20$ we find $N_\ast = 2.747$ with $\delta_2 = -3.05$ from $\beta_{\rm NL}$ 


\end{document}